\documentclass[a4paper,11pt]{article}
\pdfoutput=1 

\usepackage{jcappub}

\usepackage[T1]{fontenc}
\usepackage{afterpage}
\newcommand{\fnl}{f_{\rm NL}}

\title{CMB constraints on running non-Gaussianity}

\author[a,b]{F. Oppizzi,}
\author[a,b,c]{M. Liguori,}
\author[b]{A. Renzi,}
\author[d,e]{F. Arroja,}
\author[a,b,c]{N. Bartolo}

\affiliation[a]{Dipartimento di Fisica e Astronomia "G. Galilei", Universit{\`a} degli Studi di Padova,\\ Via Marzolo 8, 35131 Padova, Italy}
\affiliation[b]{INFN, Sezione di Padova, via Marzolo 8, I-35131, Padova, Italy}
\affiliation[c]{INAF-Osservatorio Astronomico di Padova, Vicolo dell'Osservatorio 5, I-35122 Padova, Italy}
\affiliation[d]{Leung Center for Cosmology and Particle Astrophysics, National Taiwan University, Taipei 10617, Taiwan}
\affiliation[e]{Instituto de Astrof\'{\i}sica e Ci\^{e}ncias do Espa\c{c}o, Faculdade de Ci\^{e}ncias da Universidade de Lisboa, Campo Grande, PT1749-016 Lisboa, Portugal}

\emailAdd{filippo.oppizzi@pd.infn.it}
\emailAdd{michele.liguori@pd.infn.it}
\emailAdd{alessandro.renzi@pd.infn.it}
\emailAdd{arroja@phys.ntu.edu.tw}
\emailAdd{nicola.bartolo@pd.infn.it}

\abstract{We develop a complete set of tools for CMB forecasting, simulation and estimation of primordial running bispectra, arising from a variety of curvaton and single-field (DBI) models of Inflation. We validate our pipeline using mock CMB running non-Gaussianity realizations and test it on real data by obtaining experimental constraints on the $\fnl$ running spectral index, $n_{\rm NG}$, using WMAP 9-year data.  Our final bounds (68\% C.L.) read $-0.6< n_{\rm NG}<1.4$, $-0.3< n_{\rm NG}<1.2$, $-1.1<n_{\rm NG}<0.7$ for the single-field curvaton, two-field curvaton and DBI scenarios, respectively. We show forecasts and discuss potential improvements on these bounds, using {\it Planck} and future CMB surveys.}

\begin{document}

\maketitle

\newcommand{\mmm}{m_1m_2m_3}
\newcommand{\lm}[1]{_{\ell_#1m_#1}}
\newcommand{\llll}{{\ell_1\ell_2\ell_3}}
\newcommand{\perm}[1]{+#1\,\mathrm{perm.}}

\flushbottom

\section{Introduction}

Primordial cosmological non-Gaussianity (NG) is nowadays constrained very tightly by Cosmic Microwave Background (CMB) data, 
with the most stringent bounds coming from measurements of the angular bispectrum and trispectrum of {\it Planck} temperature and polarization maps \cite{2014A&A...571A..24P,2016A&A...594A..17P}. Besides the most common "local", "equilateral" and "orthogonal" NG shapes, arising from a large variety of either single (equilateral, orthogonal) or multi-field (local) inflationary scenarios, many additional bispectra have been tested in the {\it Planck} analysis. A non-exhaustive list includes anisotropic, flattened and parity-odd shapes, as well as a large family of strongly scale-dependent, oscillatory models. However, to date, no analysis of {\it Planck} data includes scale-dependent models with a mild $\fnl$ running, described by a non-Gaussian spectral index. 

The main goal of this paper is to build a set of bispectrum estimators for a complete set of theoretically motivated running NG models. The only existing experimental constraint on running NG, to the best of our knowledge, was obtained by the authors of \cite{PhysRevLett.109.121302}. They considered a local-type running shape that is expected to appear in curvaton and modulated reheating scenarios.
In this work we will extend the analysis in various directions. First of all, we will include additional shapes, peaking both in the local and equilateral limit and generated in different multi-field inflationary models and in Dirac-Born-Infeld (DBI) scenarios. For the latter, we will implement the full, non-separable, shape reported in theoretical derivations and carefully compare it to the separable phenomenological parametrization that was employed in previous forecast analyses \cite{2009JCAP...12..022S}. Finally, before obtaining our experimental constraints, we will pay particular attention to the validation of our pipeline, via generation of NG CMB maps, including all the running bispectra under examination.
 
 The paper is organized as follows: in section \ref{sec:SDM}, we will provide a short overview of the models included in our analyses, focusing on their primordial bispectrum predictions; in section \ref{sec:method}, we will describe in detail our data analysis pipeline, including bispectrum estimation and generation of NG maps; in section \ref{sec:results} we will show the outcome of validation tests, final WMAP experimental bounds on all running shapes and forecasts for future CMB surveys; we will finally summarize our main results in section \ref{sec:conclusions}.     

\section{Scale dependent models}\label{sec:SDM}

In this work we will consider inflationary models that produce a running of the NG parameter $f_{\rm NL}$, parametrized via a NG spectral index $n_{\rm NG}$.
Running primordial NG can be sourced by a wide range of different physical processes, such as non-linear evolution of perturbations, interactions in multi-field inflation, variation of the sound speed in single-field inflationary models, peculiar properties of the background metric. Note that scale-dependence can arise also in very simple models and can therefore be considered as a fairly general prediction of Inflation.

It is actually not possible to encompass such a variety of scenarios using just a single bispectrum shape, with a specific ansatz for $f_{\rm NL}(k)$. The aim of this work is to build a quite general class of CMB bispectrum estimators, which can account for most of the theoretically motivated, scale-dependent NG parametrizations proposed so far in the literature. We start in this section by briefly reviewing them.
Scale-dependent (SD) NG in the context of slow-roll inflation was studied for example in \cite{1475-7516-2010-10-004} and \cite{2011JCAP...03..017S}, where the authors propose explicit expressions for the primordial three point function, in the cases of one or two fields contributing to the curvature perturbations.
The SD local generalization, when only one field contributes to primordial perturbations reduces to:
\begin{equation}
    B(k_1,k_2,k_3)\propto\fnl\left[(k_1k_2)^{n_\zeta -4}k_3^{n_{\rm NG}}+2\ \mathrm{perms.}\right],
    \label{eq:1f}
\end{equation}
where $n_\zeta$ denotes the usual spectral index of curvature perturbations $\zeta$. This shape was constrained using WMAP data in \cite{PhysRevLett.109.121302}. It describes multi-field models (e.g curvaton or modulated reheating) in which the Inflaton contribution to perturbations is subdominant.
A large scale dependence arises here as a consequence of strong self interactions of the field \cite{1475-7516-2010-10-004} .
In the following, we will use the superscript "\textit{1f}" to address this model.

If two fields contribute to the curvature perturbation, the dependence on $k$ follows a different parametrization. 
This kind of template arises, for example, from the mixed inflaton-curvaton theory (assuming the curvaton field has a quadratic potential), and also in general two-field models when the test field has a quadratic potential \cite{1475-7516-2010-10-004}.
The resulting shape is:
\begin{equation}
    B(k_1,k_2,k_3)\propto\fnl\left[(k_1k_2)^{n_\zeta + (n_{\rm NG}/2)-4}+2\ \mathrm{perms.}\right].
    \label{eq:2f}
\end{equation}
We will refer to this model with the superscript \textit{2f}.

Both these shapes are separable over wavenumbers, therefore it will be fairly straightforward to implement them in a generalized version of the classic KSW bispectrum estimator \cite{0004-637X-634-1-14}, as we will discuss in the next section.
In the theoretical derivation of the previous two shapes, it was assumed that the fields are slow-rolling and that $|n_\mathrm{NG}\ln \left(k_{max}/k_{min}\right)|<<1$, where $k_{max}$ and $k_{min}$ are the largest and smallest wavenumbers included in the analysis. For WMAP, $\ln \left(k_{max}/k_{min}\right)\lesssim7$ so $n_\mathrm{NG}$ can be at most of order of $0.1$. However, in this work, we wish to argue that from a purely phenomenological point of view, the previous two bispectrum shapes are interesting templates to constrain with data even when $n_{\rm NG}$ is larger than 0.1.
The value of the running, for this class of models, is proportional to higher order derivatives of the inflationary potential. Since the power spectrum is insensitive (to lowest order) to these quantities, measuring $n_{\rm NG}$ can provide additional information about primordial perturbations.
If we move to single-field scenarios with a non-canonical kinetic term, a mild running of the NG can be also produced. A typical such example is DBI-inflation \cite{PhysRevD.72.123518} (see \cite{Bartolo:2010im} for a generalization within effective field theory of Inflation). In this context, the NG amplitude $f_{\rm NL}$ is promoted to a function of the triangular wavenumbers configurations. A first study of the testability of these models is presented in \cite{1475-7516-2009-12-022}. In that work the authors proposed a parametrization assuming a dependence on the geometric mean of the three wavenumbers
\begin{equation}
B(k_1,k_2,k_3)\propto\fnl \left( \frac{k_1k_2k_3}{\textbf{k}_{piv}^3}\right)^{n_{\rm NG}/3}F(k_1,k_2,k_3) \, ,
\label{eq:gmpar}
\end{equation}
where $\textbf{k}_{piv}$ is a pivot scale and $F$ is the shape function. Being the scale independent part of the bispectrum, $F$ depends only on the ratios of the three wavenumbers.
In the theoretical literature (see e.g. \cite{Chen:2006nt}) a different parametrization is however generally found, in terms of the arithmetic, rather than the geometric mean. Namely:
\begin{equation}
B(k_1,k_2,k_3)\propto \fnl\left( \frac{k_1+k_2+k_3}{3\textbf{k}_{piv}}\right)^{n_{\rm NG}}F(k_1,k_2,k_3).
\label{eq:ampar}
\end{equation}
The geometric mean parametrization can be seen as an approximation of the theoretical shape, which is expected to work well for equilateral shapes (typical case for DBI, and in more general single-field models), where the significant contribution comes from configurations with $k_1 \sim k_2 \sim k_3$ (in that case of course, the two parametrizations coincide). Its practical advantage lies in its explicit separability. 
On the other hand, for an accurate measurement, it is important at least to compare the two parametrizations, and explicitly verify their level of correlation.
The technical problem with the arithmetic mean is that it is not explicitly separable, that is, the factor $(k_1+k_2+k_3)^{n_{\rm NG}}$ is not trivially factorizable. There are many well-known ways in the literature to circumvent this problem, based for example on the modal \cite{2010PhRvD..82b3502F,2012JCAP...12..032F} or binned \cite{Bucher:2015ura,2010MNRAS.407.2193B} decomposition of the shape.
In this work, we take a different approach. We stick with a KSW-type estimator, and factorize the shape by resorting to the so-called Schwinger parametrization \cite{Smith11102011}
\begin{equation}
B(k_1,k_2,k_3) \propto  \fnl\frac{F(k_1,k_2,k_3)}{\Gamma(1-n_{\rm NG})}\frac{1}{\textbf{k}_{piv}^{n_{\rm NG}}}\int_0^\infty\mathrm{d}t\ t^{-n_{\rm NG}}\left[k_1e^{-t(k_1+k_2+k_3)} +perm\right],
\label{eq:scpar}
\end{equation}
where $\Gamma$ is the Gamma function. This form is valid for $n_{\rm NG}<1$, which is not a very limiting assumption, since all the models predict a running $n_{\rm NG} \lesssim10^{-1}$. We will refer to this shape with the superscript {\it``am''}. For an overview of the explicit form of the resulting CMB templates, see Appendix \ref{ap:templates}. The advantage of this approach, in this specific case, is that we will not have to re-expand the shape every time we change $n_{\rm NG}$, while exploring the parameter space.  

\section{Methodology}\label{sec:method}

We aim to extend the KSW NG estimation technique \cite{0004-637X-634-1-14} (for a comprehensive review about CMB NG estimation see e.g. \cite{2010AdAst2010E..73L}, and references therein) to include all types of bispectra described in the previous section. We follow the indirect approach developed in \cite{PhysRevLett.109.121302}: we first apply an estimator of the NG amplitude $f_{\rm NL}$, for different, fixed values of the running $n_{\rm NG}$. With this set of measurements we reconstruct the full likelihood function  $\mathcal{L}(\hat{f}_{\rm NL},n_{\rm NG})$, and finally obtain a joint fit of both parameters. A brief description of the standard KSW $f_{\rm NL}$ estimator is reported in Appendix \ref{app:KSW}. We review here its extension to include 
the running parameter $n_{\rm NG}$. We will follow the approach firstly introduced in \cite{PhysRevLett.109.121302}, introducing some modifications, that will be discussed in the following.
\begin{figure}
    \centering
    \includegraphics[width=1.\textwidth]{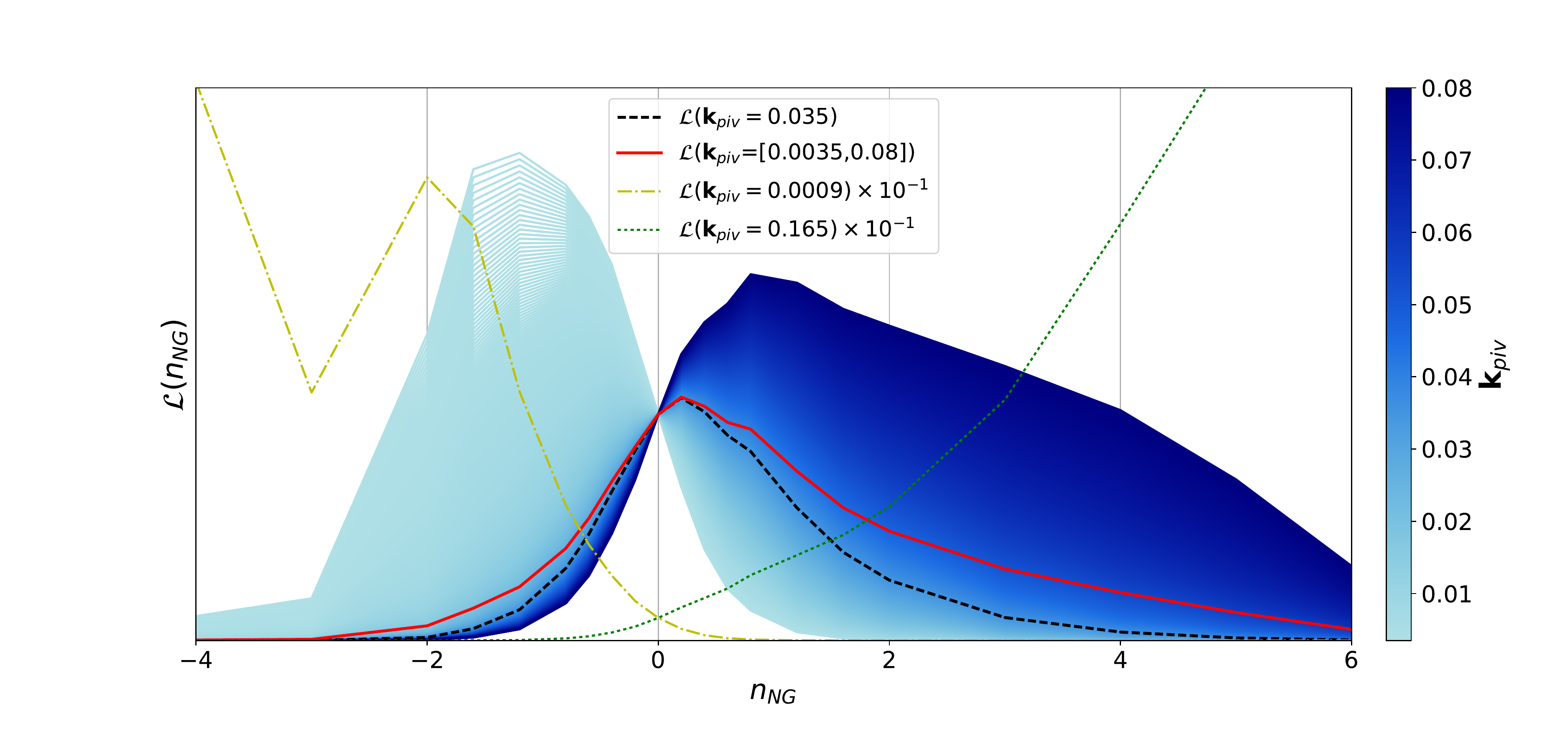}
    \caption{From light blue to dark blue: likelihood functions for increasing values of the pivot scale, in the range $[0.0035, 0.08] \, \mathrm{Mpc}^{-1} $. Dashed black line: likelihood for the best fit $\mathbf{k}_{piv}$ (see the text). Continous red line: likelihood averaged over $\mathbf{k}_{piv}$ in the same interval. Dash-dotted yellow line: likelihood for $\mathbf{k}_{piv} = 9 \times 10^{-4} \, \mathrm{Mpc}^{-1}$ rescaled by a factor $10^{-1}$. Dotted green line: likelihood for $\mathbf{k}_{piv} = 0.165 \, \mathrm{Mpc}^{-1}$ rescaled by a factor $10^{-1}$. }
    \label{fig:kpivcomp}
\end{figure}
 Under the assumption of normally distributed bispectrum configurations, the likelihood can be written as:
\begin{equation}
\mathcal{L}(f_{\rm NL},n_{\rm NG})\propto \exp\left[-\frac{1}{2}\sum_\llll\frac{ \left( B_\llll^{obs}-f_{\rm NL}B_\llll^{th}(f_{\rm NL}=1,n_{\rm NG})\right)^2}{C_{\ell_1}C_{\ell_2}C_{\ell_3}}\right].
\end{equation}
Dropping constant terms, this equation can be rearranged in the form:
\begin{equation}
\mathcal{L}(f_{\rm NL},n_{\rm NG})\propto\exp\left[ -\frac{f_{\rm NL}^2}{2}\sum_\llll 
\frac{\left(B_\llll^{th}\right)^2}{C_{\ell_1}C_{\ell_2}C_{\ell_3}}+f_{\rm NL}\sum_\llll \frac{B_\llll^{obs}B_\llll^{th}}{C_{\ell_1}C_{\ell_2}C_{\ell_3}} \right].
\end{equation}
It is easy to recognize, in the first term, the KSW normalization factor $\mathcal{N}$ (see formula \ref{eq:norm}) and, in the second term, the unnormalized estimator (see formula \ref{eq:kswcub}). We can therefore re-write the expression above as:
\begin{equation}
\mathcal{L}(f_{\rm NL},n_{\rm NG})\propto\exp\left[\mathcal{N}\left(-\frac{f_{\rm NL}^2}{2}+f_{\rm NL}\hat{f}_{\rm NL}\right)\right],
\end{equation}
where $\hat{f}_{\rm NL}$ is the value of the NG amplitude recovered from the KSW estimator.
Assuming a constant prior on $\fnl$, we can integrate to find the marginalized likelihood of $n_{\rm NG}$:
\begin{equation}
\mathcal{L}(n_{\rm NG})\propto\frac{\mathbf{k}_{piv}^{n_{\rm NG}}}{\sqrt{N}}\exp\left(\frac{\hat{f}^2_{NL}\mathcal{N}}{2} \right),
\end{equation}
here we have explicitly highlighted the dependence of the likelihood on the pivot scale $\mathbf{k}_{piv}$, and $N$ denotes the normalization without the pivot factor.
Due to the limitations in resolution, the experimental sensitivity is not constant for different choices of different $\mathbf{k}_{piv}$. As a consequence, the correlation between the two parameters depends on the pivot scale, and this reflects on the shape of the marginalized likelihood.
An example of the dependence of $\mathcal{L}(n_{\rm NG})$ on $\mathbf{k}_{piv}$ is shown in figure \ref{fig:kpivcomp}, where we consider the one-field model likelihood obtained from a Gaussian simulation
(as extreme examples, we show also cases in which we set the pivot outside the accessible scales, resulting in a likelihood that diverges at the edges).  
The standard approach, (see e.g. \cite{1475-7516-2009-12-022}\cite{PhysRevLett.109.121302}), which we also follow here,  is to start with an arbitrary value of $\mathbf{k}_{piv}$, compute the likelihood and finally rescale  $\mathbf{k}_{piv}$ in order to minimize the correlation between the parameters at the peak of the likelihood. 
\begin{figure}
    \centering
    \includegraphics[width=1.\textwidth]{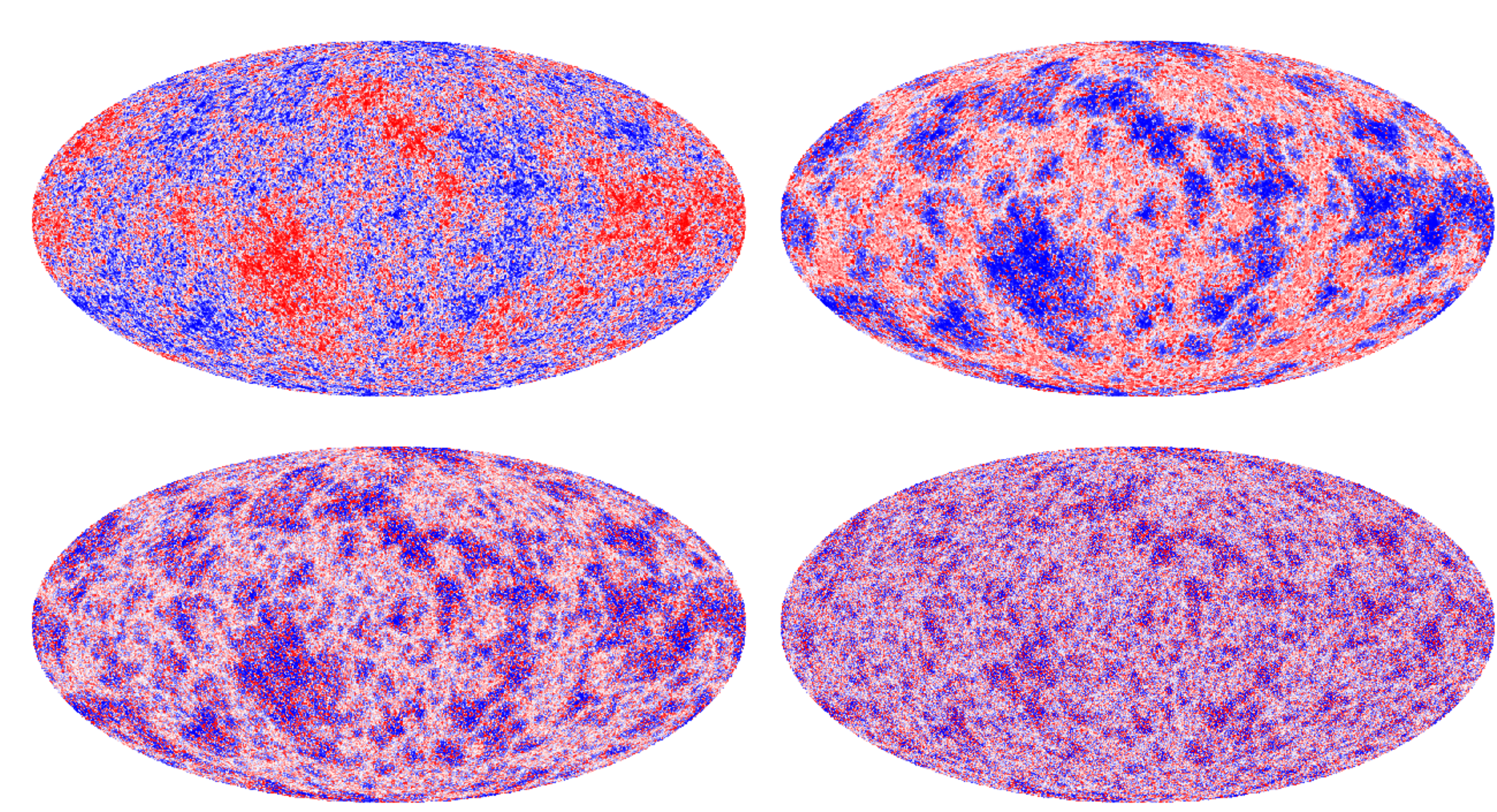}
    \caption{ Random realizations of CMB sky with a SD bispectrum from model \ref{eq:1f} for different values of the parameters. Upper left panel: Gaussian field ($\fnl=0$). Upper right, lower left and lower right panels: non-Gaussian residuals (i.e. residuals after the subtraction of the Gaussian  component of the field) for simulations with, respectively $n_{\rm NG}=-0.2$, $n_{\rm NG}=0.2$, $n_{\rm NG}=0.8$ .}
    \label{fig:ngsym}
\end{figure}
As mentioned above, the estimated amplitude $\hat{f}_{\rm NL}(n_{\rm NG})$ is computed via KSW bispectrum estimation, for fixed $n_{\rm NG}$. The likelihood is then profiled by iterating this operation over a sufficiently wide $n_{\rm NG}$ interval. Finally, with the full profiled likelihood in hand, we can extract the best fit value of $n_{\rm NG}$.
In case of partial sky coverage and non-stationary noise, it is well known that a linear term must be added to the KSW cubic statistic, to restore optimality. This is generally computed as a Monte Carlo average over Gaussian realizations of the masked CMB sky, including realistic instrumental noise properties. 
When producing these realizations, for the WMAP analysis that follows, we adopt the procedure described in \cite{2009ApJS..180..330K}, adding to it an inpainting procedure of masked regions, as done in {\it Planck} data analysis \cite{2014A&A...571A..24P}. For a complete discussion on inpainting and for studies of its efficiency in the context of CMB bispectrum estimation see \cite{Gruetjen:2015sta,Bucher:2015ura}.

\subsection{Simulations of non-Gaussian maps}

\begin{figure}
    \centering
    \includegraphics[width=1.\textwidth]{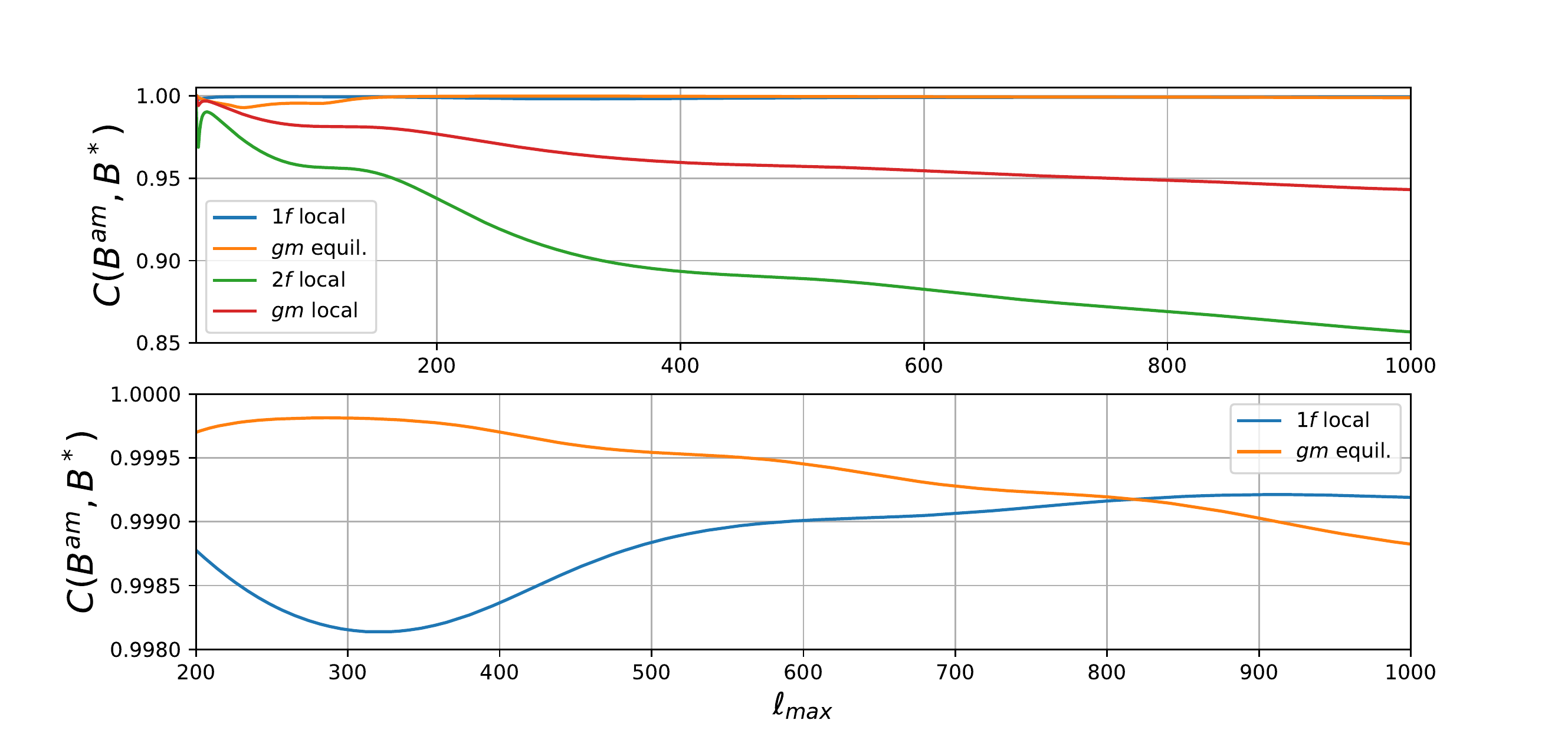}
    \caption{Correlation between shapes as a fuction of the maximum multipole number, computed with respect to the arithmetic mean parametrization of the respective template (local for one-field, two-fields and the geometric mean local, equilateral for geometric mean equilateral). All correlations are computed for $n_{\rm NG}=0.8$. The lower panel shows a zoom on the templates with higher correlation.}
    \label{fig:corr}
\end{figure}
In order to test our estimators, we produce CMB simulations with a scale dependent $f_{\rm NL}$.
We follow the method described in \cite{Smith11102011}, which is valid for any separable primordial shape. We can therefore apply it in all the cases under study, since we are always able to find separable representations for our models, eventually via Schwinger parametrization in the technically most complex case.
We can express the CMB multipoles as a sum of a Gaussian and a non-Gaussian part, so that:
\begin{equation}
    a_{\ell m}=a_{\ell m}^{G}+f_{\rm NL}a_{\ell m}^{NG},
    \label{eq:multng}
\end{equation}
where the non-Gaussian component is expressed as:
\begin{equation}
a_{\ell m}^{NG}=\frac{1}{6}\sum_{\ell_1 \ell_2}^{m_1 m_2}B_{\ell\ell_1\ell_2}^{th}
\begin{pmatrix} 
\ell & \ell_1 & \ell_2 \\
m & m_1 & m_2
\end{pmatrix}
\frac{a_{\ell_1 m_1}^{G}}{C_{\ell_1}}\frac{a_{\ell_2 m_2}^{G}}{C_{\ell_2}}.
\end{equation}
We compute the NG multipoles inserting in this equation the theoretical separable bispectrum template under study (see appendix \ref{ap:templates} for the expressions).
Once computed the $a_{\ell m}^{NG}$, it is straightforward to obtain the final map from Eq.  $\ref{eq:multng}$ for any value of $f_{\rm NL}$.
This methods, holding in the limit of small NG, 
has a drawback in that it can fail in controlling terms $\mathcal{O}(f_{\rm NL}^2)$ in connected N-point functions \cite{2009PhRvD..80h3004H}.
This can bring a significant, spurious contribution to the power spectrum from the non-Gaussian component.
A technique to control and remove this spurious signal has been developed and is described in depth in \cite{2010PhRvD..82b3502F}, to which we refer for further details about NG simulations.

In Figure \ref{fig:ngsym} we show the contributions on the CMB sky of a SD bispectrum from model \ref{eq:1f} for different values of the running. All maps are computed starting from the same Gaussian multipoles. To highlight the differences, we plot the Gaussian and the non-Gaussian components separately. It is evident from these maps how the non-Gaussian modulation peaks on smaller scales for higher values of the running.

\subsection{Correlation between shapes}\label{sec:corr}
\begin{figure}
    \includegraphics[width=0.49635\textwidth]{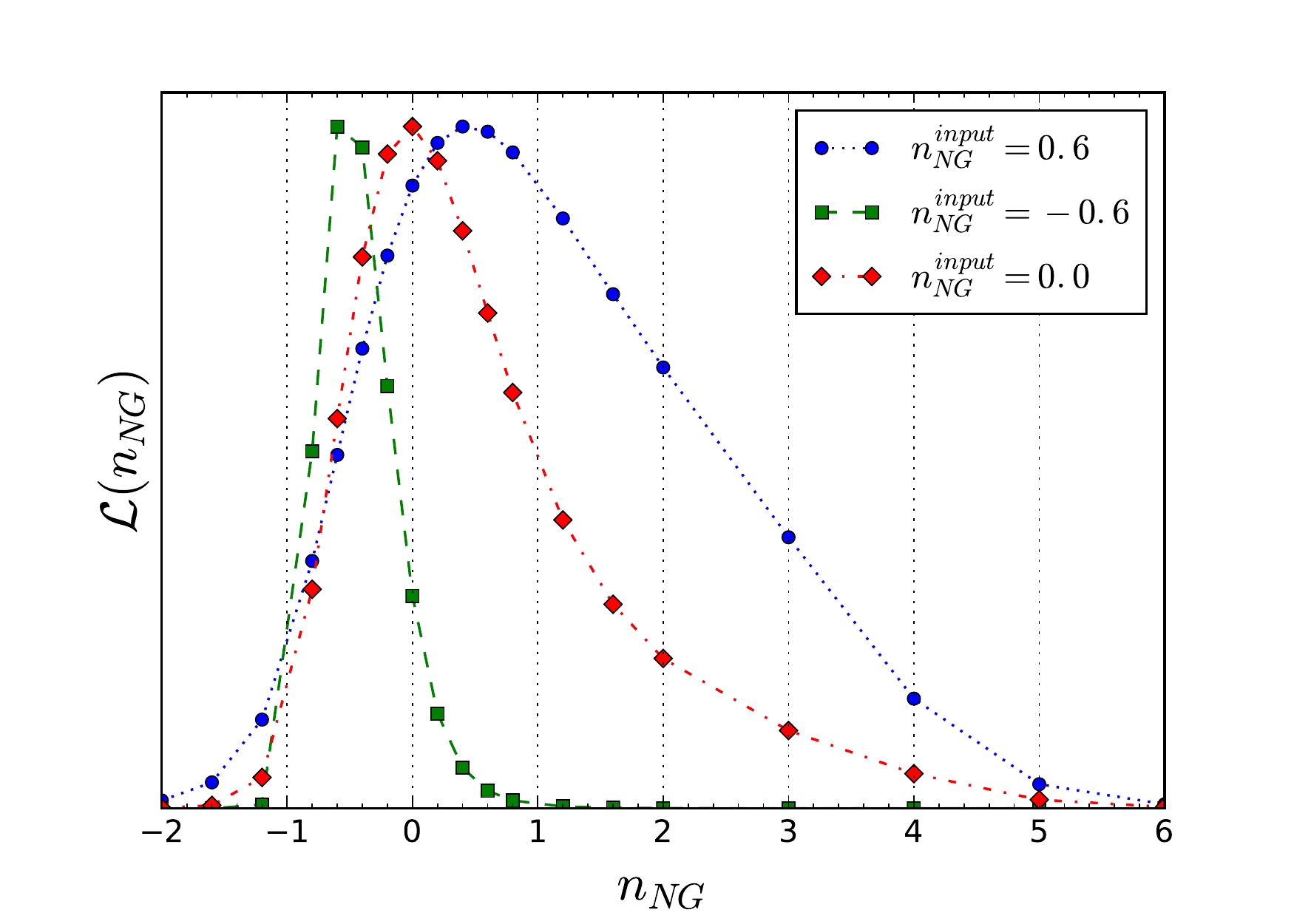}
    \includegraphics[width=0.49635\textwidth]{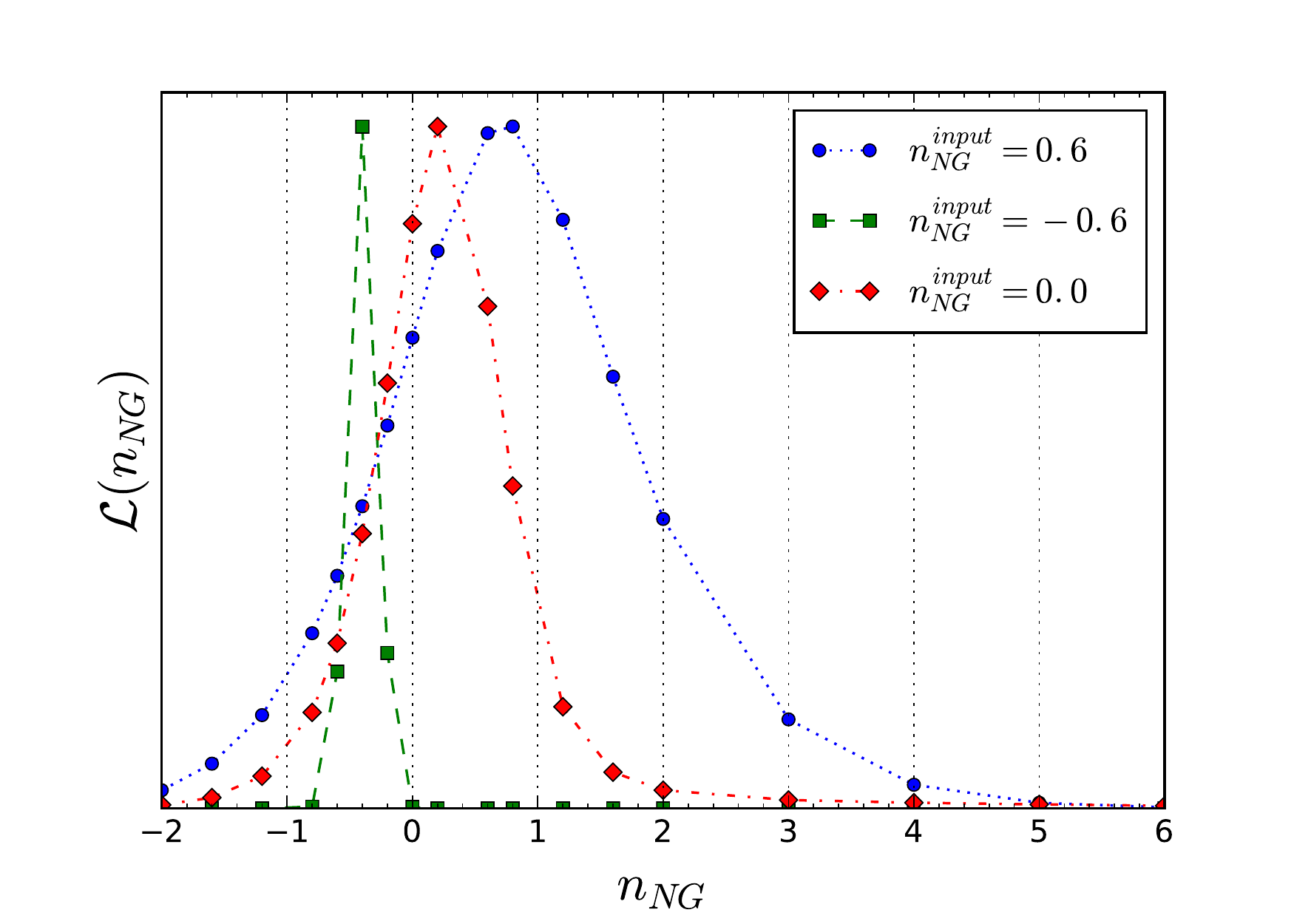}
    \caption{Likelihood from the one-field (left) and two-field (right) models, for different input value of $n_{\rm NG}$.}\label{fig:comp}
\end{figure}
We compute the correlation between the shapes described before.
The correlation is obviously higher for low values of $n_{\rm NG}$, since all templates reduce to the classical scale invariant case in the limit of zero running.
We define the correlation between shapes as the portion of the total NG amplitude recovered fitting a template $B'$ if the true bispectrum is $B$.
The expression of the correlator is:
\begin{equation}
C(B,B')=\frac{1}{N(B,B')}\sum_\llll\frac{B_\llll B'_\llll}{C_{\ell_1}C_{\ell_2}C_{\ell_3}},
\end{equation}
where the normalization $N(B,B')$ is:
\begin{equation}
N=\sqrt{\sum_\llll\frac{ì(B_\llll)^2}{C_{\ell_1}C_{\ell_2}C_{\ell_3}}}\sqrt{\sum_\llll\frac{(B'_\llll)^2}{C_{\ell_1}C_{\ell_2}C_{\ell_3}}}.
\end{equation}
This analysis clearly shows that some of the templates under study are highly correlated. As expected, in the equilateral case, the arithmetic mean and the geometric mean parametrizations are almost equivalent, having $C(B^{am},B^{gm})\sim0.999$. In the local case, an high level of correlation is instead displayed between the arithmetic mean parametrization and the one-field model \ref{eq:1f}, where $C>99.9\%$ for all the multipoles of interest. This behavior is highlighted in figure \ref{fig:corr}, in which we show correlations between the templates, as a function of the maximum multipole considered. For the one-field, two-fields and geometric mean local parametrization the correlations are computed with respect the arithmetic mean parametrization of the local shape. For the geometric mean equilateral shape we show the correlation between the arithmetic mean equilateral parametrization. To better highlight differences between templates, we consider an extreme value for the running, $n_{\rm NG}=0.8$. It is also interesting to note how the two-field model \ref{eq:2f} is highly uncorrelated with the others, showing the importance of separately fitting the different scale-dependent shapes predicted in the literature.

\section{Results}\label{sec:results}
%

\subsection{Test on simulations}
%

To test our estimator, we run it on different sets of NG maps, produced with the method outlined in the previous section.
We produce nine different sets, with different spectral index, for three models: the one-field and two-field local models, and the equilateral "geometric mean" model.
We choose a value $f_{\rm NL}=50$ for the local templates and $f_{\rm NL} = 100$ for the equilateral one, at a pivot scale $\mathbf{k}_{piv}=0.02 \ \mathrm{Mpc}^{-1}$.
For each model we consider three different values of the running: $n_{\rm NG}=0$, $n_{\rm NG}=-0.6$ and $n_{\rm NG}=0.6$.
We compute the Gaussian component assuming the best-fit {\it Planck} power spectrum. 
The angular resolution in these test maps is $\ell_{max}=500$.
We test our method both in the case of full sky-coverage and in a more realistic case with $30\%$ of the sky masked.

We find that, in all cases, our estimators recover correctly the initial value of the parameters, within error bars. At the same time, the uncertainties derived from the likelihood are consistent with Fisher matrix predictions. We find that $100$ Gaussian simulations are sufficient in the linear term evaluation, to correct for the partial sky coverage effects.
As an example, in figure \ref{fig:comp} we show the likelihoods obtained for the one and two-field local models from  simulated NG maps with different value of the spectral index.

\subsection{Experimental bounds}
\afterpage{
\begin{figure}
    \includegraphics[width=0.325\textwidth]{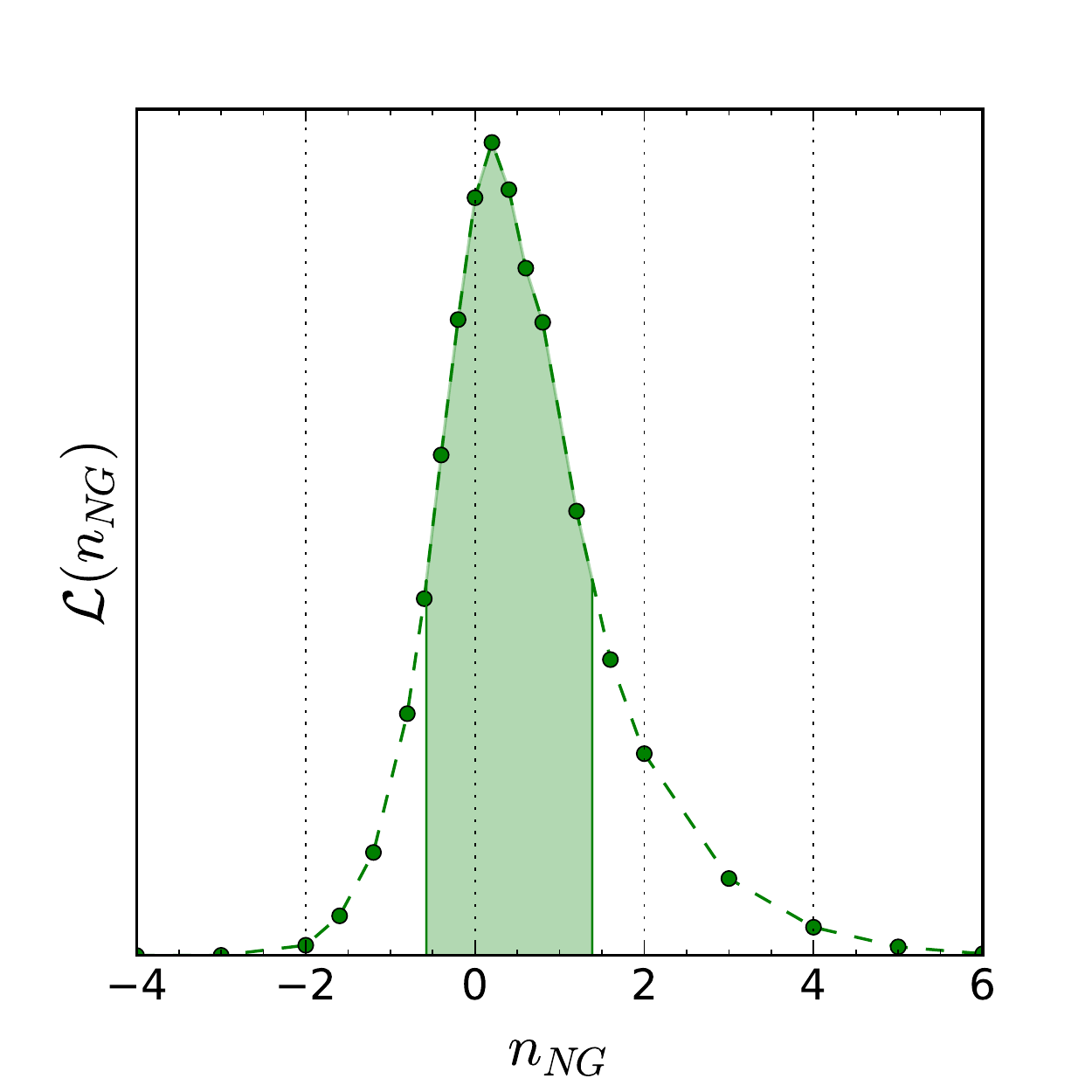}
    \includegraphics[width=0.325\textwidth]{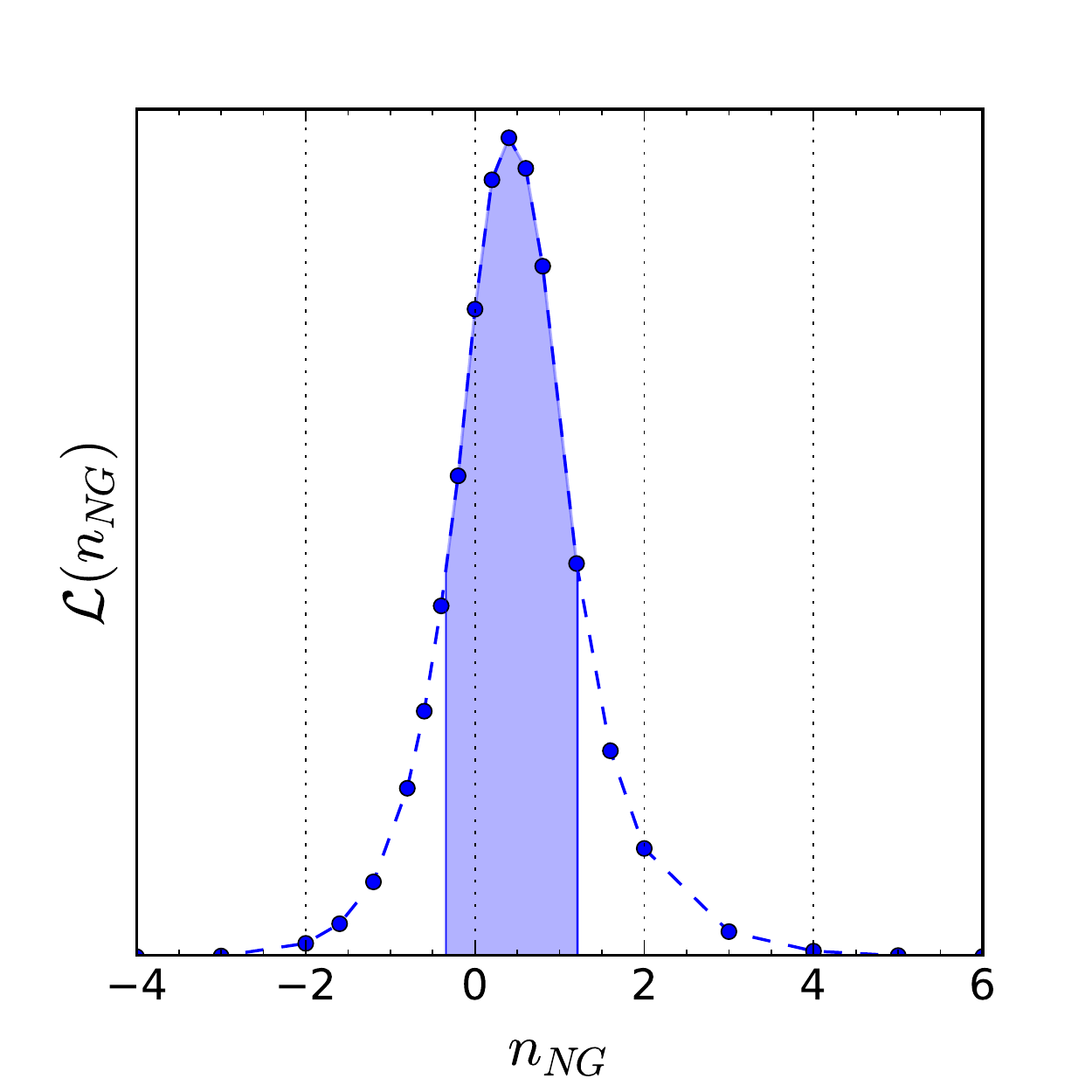}
    \includegraphics[width=0.325\textwidth]{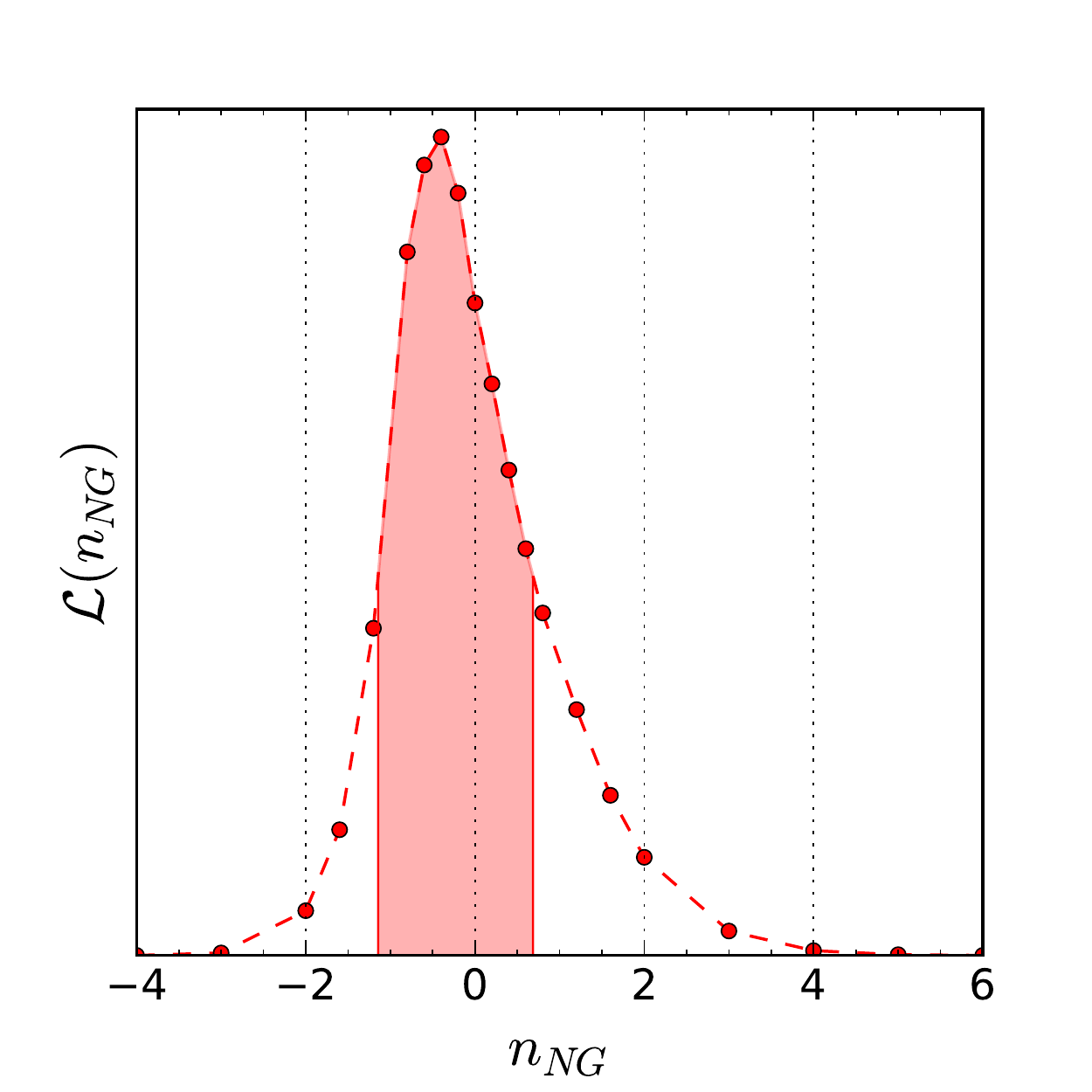}
    \caption{Marginalized likelihood for the one-field (left, green), two-field (center, blue) models and the geometric mean parametrization for the equilateral model (right, red). The shaded regions represent the $1\sigma$ intervals and were found by slicing at constant marginalized likelihood.}
    \label{fig:like}
\end{figure}
\begingroup
\renewcommand*{\arraystretch}{1.5}
\begin{table}
    \centering
    \begin{tabular}{|c|c|c|}
    \hline  \textbf{model}  & $n_{\rm NG}$ & $\mathbf{k}_{piv}$ \\
    \hline  one-field (local)        & $0.2^{+1.2}_{-0.8}$ & $0.035 \, \mathrm{Mpc}^{-1}$ \\
    \hline  two-field (local)       & $0.4^{+0.8}_{-0.7}$ &$0.01 \, \mathrm{Mpc}^{-1} $ \\
    \hline  geometric mean (equil.)  & $-0.4^{+1.1}_{-0.7}$ & $0.01 \, \mathrm{Mpc}^{-1}$ \\
    \hline  
    \end{tabular}
    \caption{Experimental constraints for the different models considered in this work. First column: model. Second column: central values and error bars (68\% C.L.). Third column: Pivot scale (see section \ref{sec:method} for details on the choice of $\mathbf{k}_{piv}$).}
    \label{tab:res}
\end{table}
\endgroup
}

We apply our technique to provide bounds on the running of non-Gaussianity from the WMAP9 temperature maps, considering the various running models discussed in the previous sections.
Our data-set consists on the combination of the V and W WMAP bands, coadded and weighted following the prescription in \cite{2009ApJS..180..330K}. 
To exclude foreground and point-source contaminated pixels we use the KQ75y9 mask, covering $31.2\%$ of the sky.
The maps used in this analysis, as well as instrumental specifications, beams and noise per pixel are extensively described in \cite{2013ApJS..208...20B} and fully available at \url{https://lambda.gsfc.nasa.gov}.
We obtain our estimates using multipoles up to $\ell_{max}=800$. 
To compute the linear correction term (\ref{eq:kswlin}) we use 300 Gaussian realizations obtained with the procedure described in section \ref{sec:method}, assuming WMAP9 cosmology and instrumental specifications. 

To further improve the numerical efficiency of our code, we implement the numerical optimization algorithm presented in \cite{Smith11102011}. This technique considerably reduces the number of terms needed in the position-space computations of the bispectra. For most models, we start from a conservative number of quadrature points $\mathcal{O}(10^3)$ for the computation of the $r$ integral in \ref{eq:kswcub}, and reduce it to $\mathcal{O}(10)$ after the optimization. For the arithmetic mean parametrization, which contains an additional integral over the Schwinger parameter $t$, we reduce the number of terms from $\mathcal{O}(10^4)$ to $\mathcal{O}(10^2)$.

The resulting likelihoods for all models are provided in figure \ref{fig:like} and the corresponding bounds are summarized in table \ref{tab:res}. The one-field model (figure \ref{fig:like}, left) is the only one that was constrained before, using CMB WMAP 7-year  data \cite{PhysRevLett.109.121302}. In that work, the authors found $n_{\rm NG}=0.30^{+0.78}_{-0.61}$. This is well consistent with our findings, which however lead to a somewhat larger confidence interval. The main source of difference is the shift in $f_{\rm NL}$ central value, which becomes a bit lower, going from WMAP 7-year to WMAP 9-year data and including inpainting at higher resolution. This lowers the sensitivity to $n_{\rm NG}$ (see also the next section for further discussion on this issue).

\begin{figure}
    \includegraphics[width=0.49635\textwidth]{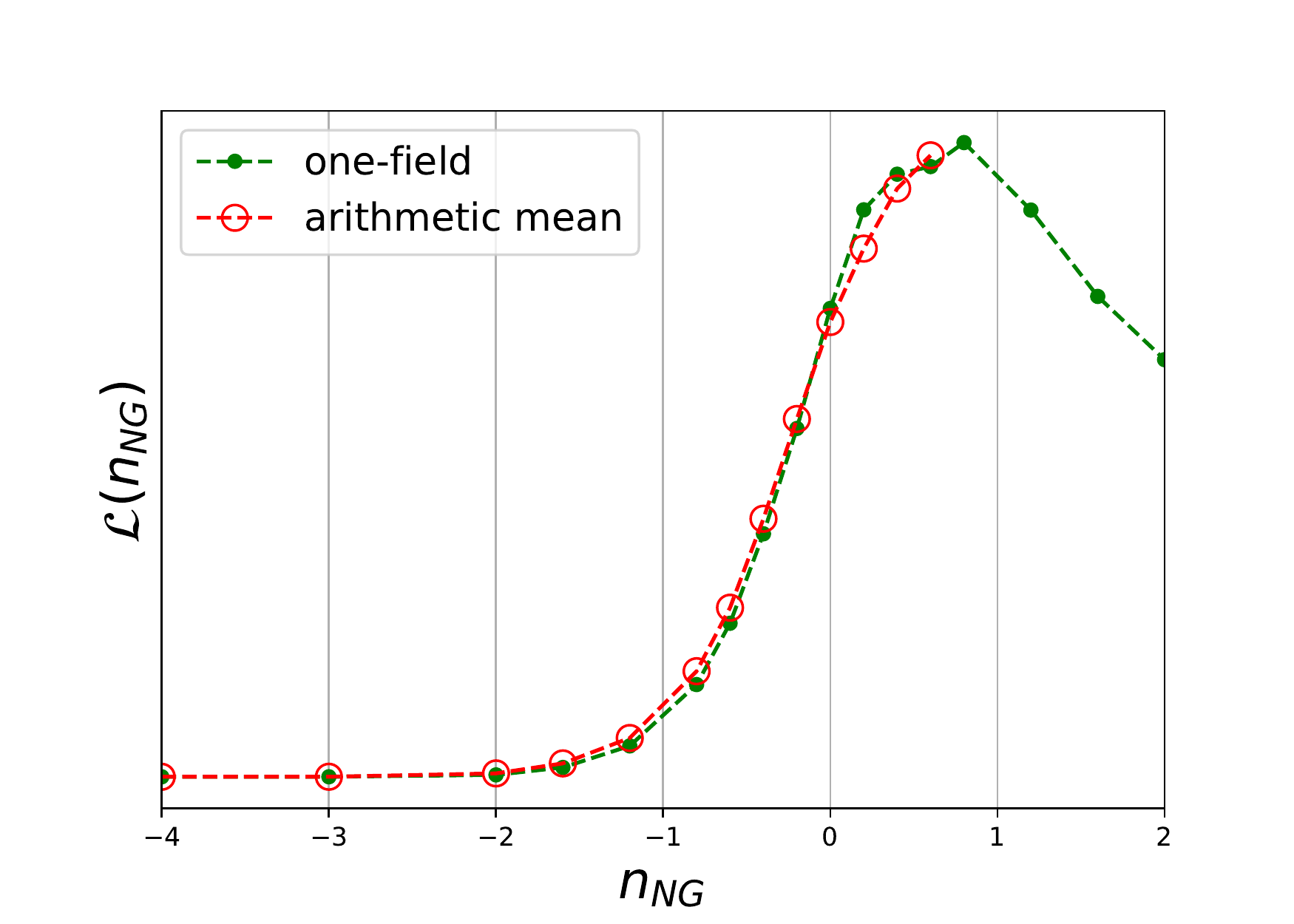}
    \includegraphics[width=0.49635\textwidth]{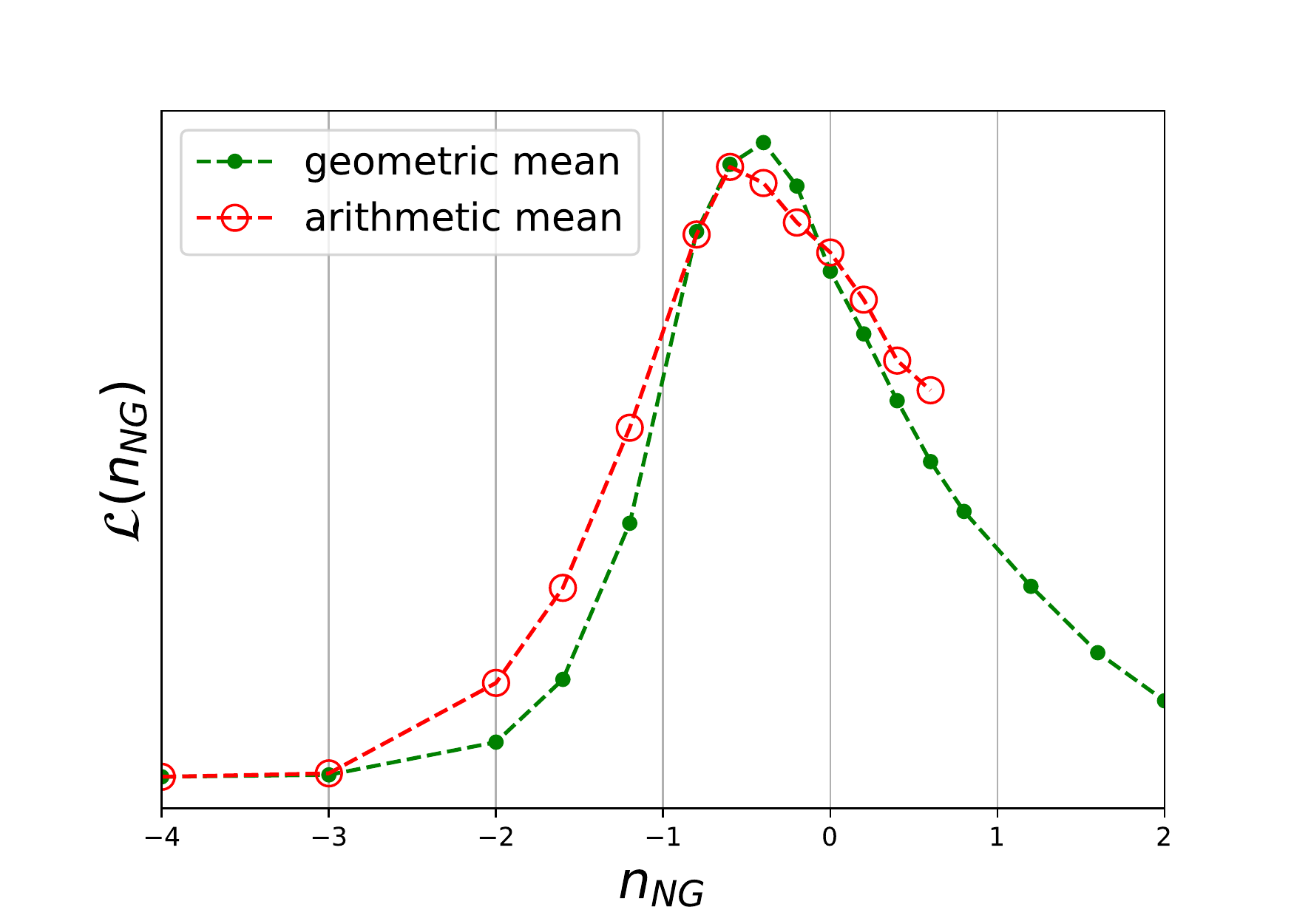}
    \caption{Comparison between the results from the arithmetic mean parametrization and the corresponding model for local (left) and equilateral (right). Left: green line: one-field local model, red line: arithmetic mean local model. Right: green line: geometric mean equilateral model red line: arithmetic mean equilateral model.}
        \label{fig:comp2}
\end{figure}

In order to obtain the full likelihood, we have in principle to span over the full $n_{\rm NG}$ parameter space, going outside the allowed domain of application of the Schwinger parametrization. Therefore, the arithmetic mean model \ref{eq:ampar} cannot be implemented everywhere. 
We considered modifications to extend the regime of validity of this parametrization, e.g. using inverse Laplace transform instead of Schwinger parametrization or, more trivially, expanding the non-separable term via multinomial expansion. However these techniques turn out to be too numerical inefficient and unstable for our purposes. 
However, this is not a big problem. We have in fact explicitly checked (see figure \ref{fig:corr}) that the arithmetic mean template becomes heavily correlated to other models under study: the one-field shape in the local case, and the geometric mean parametrization for the equilateral (note that a reasonably high level of correlation was somewhat assumed, based on general arguments, but not explicitly evaluated and checked in previous works, such as \cite{2009JCAP...12..022S}).
Given current experimental sensitivity and such levels of correlation, fitting the one-field and geometric mean separable templates is therefore equivalent (i.e. leads to the same bounds) to explicitly estimating the arithmetic mean parameterization. This would no longer be the case for very high precision measurements, namely for sensitivities of the order of $\Delta n_{\rm NG} \sim 0.1$. We do not expect such sensitivities to be however achievable, neither with {\it Planck}, nor with future proposed surveys, due to the very small values of $f_{\rm NL}$ currently constrained by data. 
As a further verification of this, we show in figure \ref{fig:comp2} the likelihood points from the arithmetic mean model, compared to the two correlated shapes in the equilateral and local cases, over the $n_{\rm NG} < 1$ range, accessible to all templates.

\subsection{Forecasts}
\afterpage{
\begin{figure}[t]
    \includegraphics[width=0.49635\textwidth]{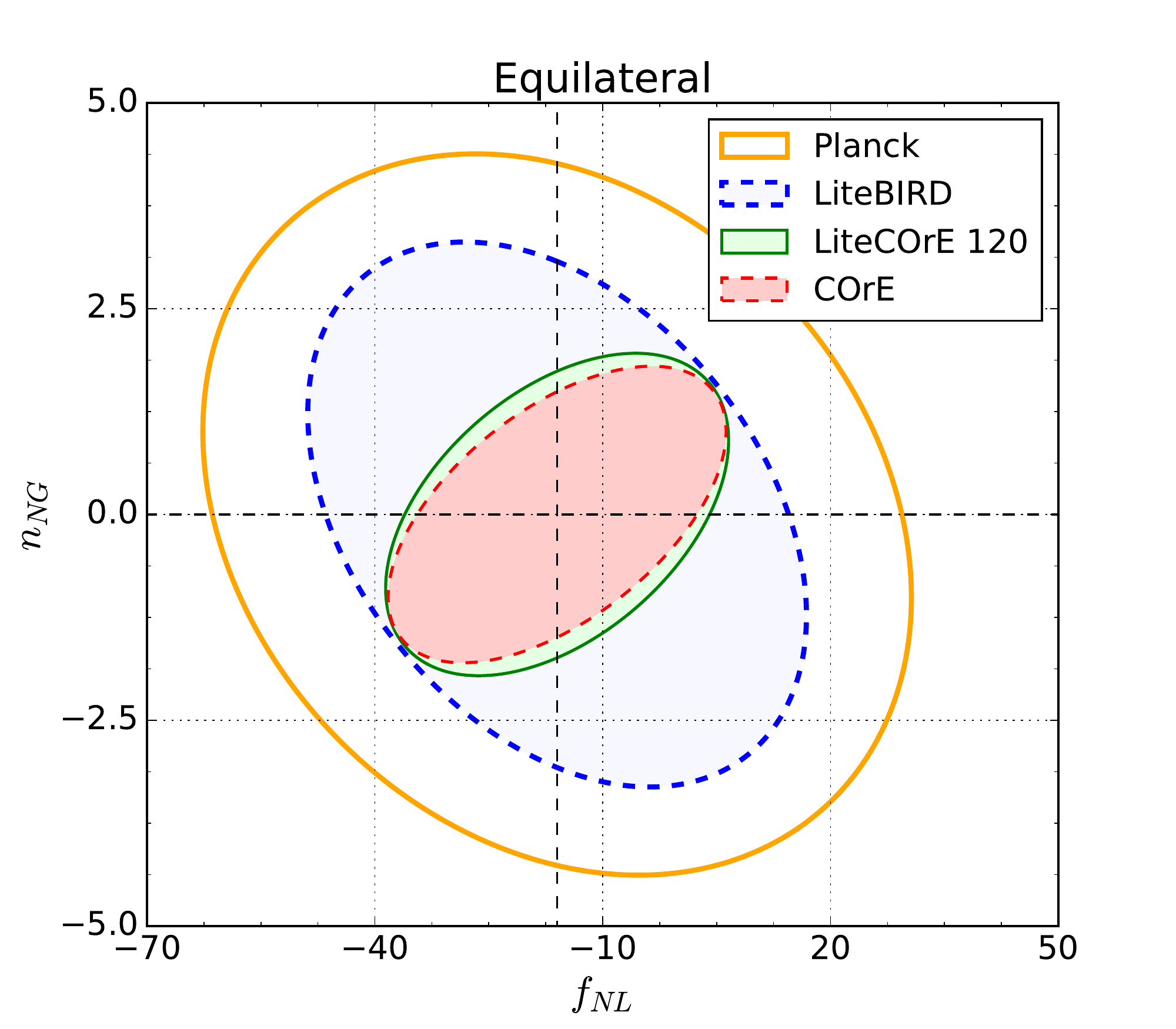}
    \includegraphics[width=0.49635\textwidth]{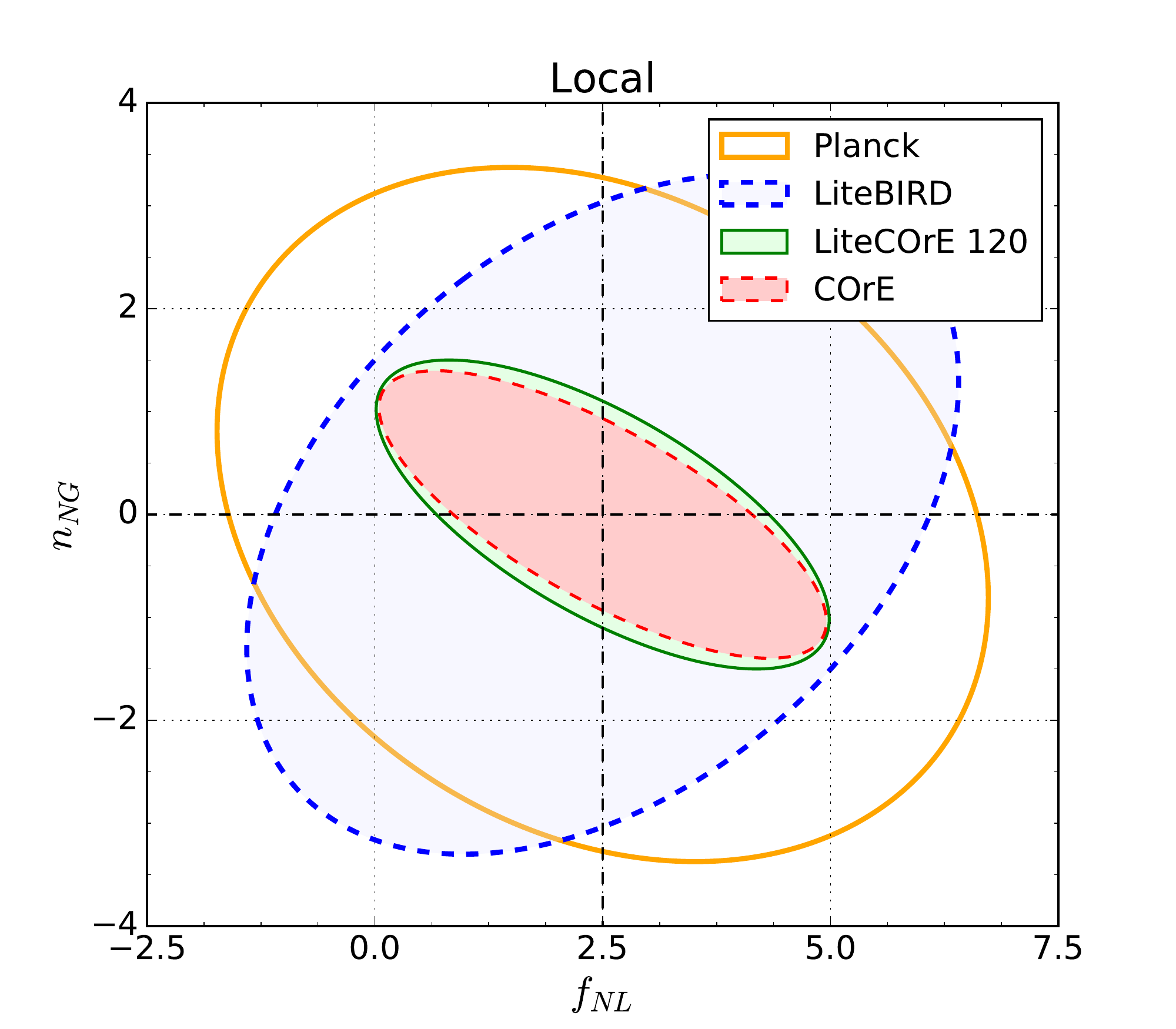}
    \caption{$1\sigma$ error ellipses in the $f_{\rm NL}-n_{\rm NG}$ plane for equilateral (left) and local (right) models arithmetic mean parametrization. We consider a joint temperature and polarization E modes analysis. We assume as central value $f_{\rm NL}^{loc}=2.5$ and $f_{\rm NL}^{eq}=-16$, $n_{\rm NG}=0$, $\mathbf{k}_{piv}=0.055\,\mathrm{Mpc}^{-1}$. Note that the orientation of the ellipses depends on the sign of $\fnl$. We find here positive correlation, for equilateral shapes, between $\fnl$ and $n_{\rm NG}$ in higher resolution experiments, because we chose a negative fiducial value for $\fnl$.}
    \label{fig:fore}
\end{figure}
\begin{table}
    \centering
    \begin{tabular}{|c c|c|c|}
    \hline  \textbf{Experiment} &($\ell_{max}$) & \textbf{local} & {\bf equilateral} \\
    \hline  Planck &(2400) & $3.4$ & $4.4$ \\
    \hline  LiteBIRD &(1350) & $3.3$ & $3.4$ \\
    \hline  LiteCOrE 120 &(3000) & $1.5$ & $1.9$ \\
    \hline  COrE &(3000) & $1.4$ & $1.8$ \\
    \hline
    \end{tabular}
    \caption{Forecasts for the marginalized $1\sigma$ $n_{\rm NG}$ error bars for the arithmetic mean parametrization, assuming joint temperature and polarization analysis. We take as central values $f_{\rm NL}^{loc}=2.5$ and $f_{\rm NL}^{eq}=-16$, $n_{\rm NG}=0$, $\mathbf{k}_{piv}=0.055\,\mathrm{Mpc}^{-1}.$  }
    \label{tab:fore}
\end{table}
}
As a final application, we forecast the combined sensitivity to the NG amplitude $\fnl$ and the spectral index $n_{\rm NG}$ of the Planck satellite and other proposed CMB projects: LiteBIRD and COrE, in various configurations. For the actual specification used for the various experiments, we refer to \cite{2016arXiv161208270C} (some $n_{\rm NG}$ forecasts, obtained with the pipeline presented here, where actually already included in \cite{2016arXiv161208270C}).
Figure \ref{fig:fore} represents the $1\sigma$ ellipses in the $\fnl - n_{\rm NG}$ plane, for the local and equilateral cases assuming the arithmetic mean parametrization.
In this computation we have assumed as fiducial values $n_{\rm NG}=0$, $f_{\rm NL}^{loc}=2.5$, $f_{\rm NL}^{eq}=-16$ at a pivot scale of $\mathbf{k}_{piv}=0.055\, \mathrm{Mpc}^{-1}$. 
For all experiments we have considered a sky coverage of $70\%$. 
The maximum multipole order accessible is: $\ell_{max}=2400$ for {\it Planck}, $\ell_{max}=1350$ for LiteBIRD, and $\ell_{max}=3000$ for the other experiments.
While these surveys are clearly signal-dominated to much higher $\ell$ than WMAP, it is not obvious that they will provide significant better constraints on $n_{\rm NG}$. That is because the uncertainty on $n_{\rm NG}$ is inversely proportional not only to $\ell_{\rm max}$ but also to the NG amplitude parameter, $\fnl$. The final forecast is therefore crucially dependent on the fiducial value chosen for $\fnl$. If we consider the current local scale-independent $\fnl$ central value from {\it Planck} analysis, we know that it has moved much closer to $0$, with respect to WMAP, making forecasted {\it Planck} constraints actually weaker than what we obtained here. A similar reasoning applies also for the equilateral shape, although the shift in measured central values from WMAP to {\it Planck} is less dramatic there.    
On the other hand, it is clear that changing the central value of e.g. local $\fnl$, in a scale-independent analysis, from approximately $30$ at $\ell_{\rm max} = 500$, to $\fnl = 2.5$ at $\ell_{\rm max} = 2000$, does display, a posteriori, some degree of running. Allowing for a further running parameter can therefore lead to a shifting of the overall $\fnl$ amplitude to larger values. Moreover, for values of $\fnl$ which do not correspond to the current best-fit value but are well within the current scale-independent 95\% C.L. intervals, significant $n_{\rm NG}$ improvements are expected with future surveys. This is evident from the results shown in table \ref{tab:forediffnl} and figure \ref{fig:forsca}. In summary, while there is a possibility that the constraints obtained here will not be significantly improved with {\it Planck} or future CMB data, several plausible scenarios do allow for significant tightening of current error bars, up to factors of $2$-$3$. This makes further studies, using more sensitive, higher resolution than WMAP datasets, clearly worth pursuing. 

\section{Conclusions}\label{sec:conclusions}
\afterpage{
\begin{figure}
    \includegraphics[width=0.325\textwidth]{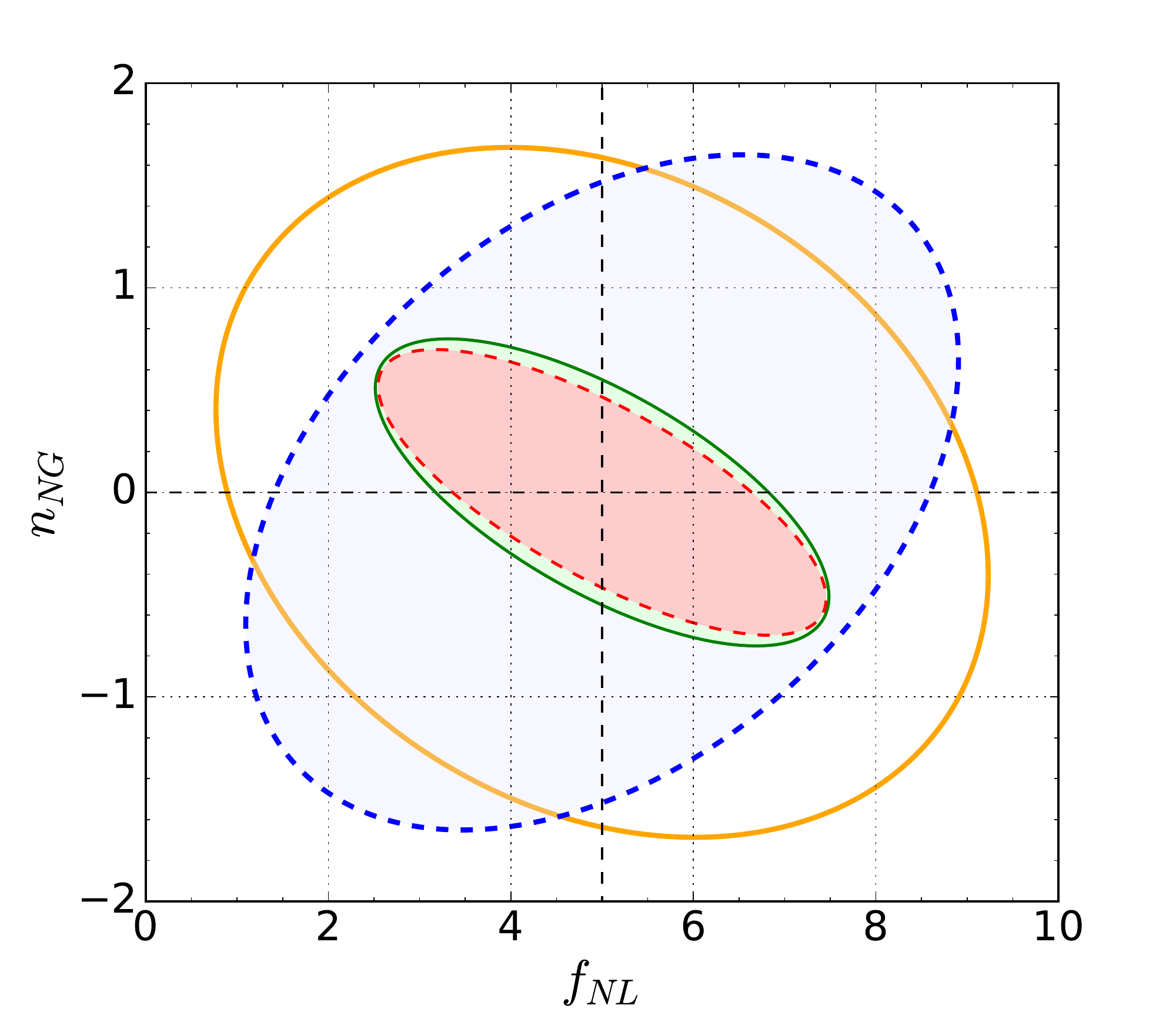}
    \includegraphics[width=0.325\textwidth]{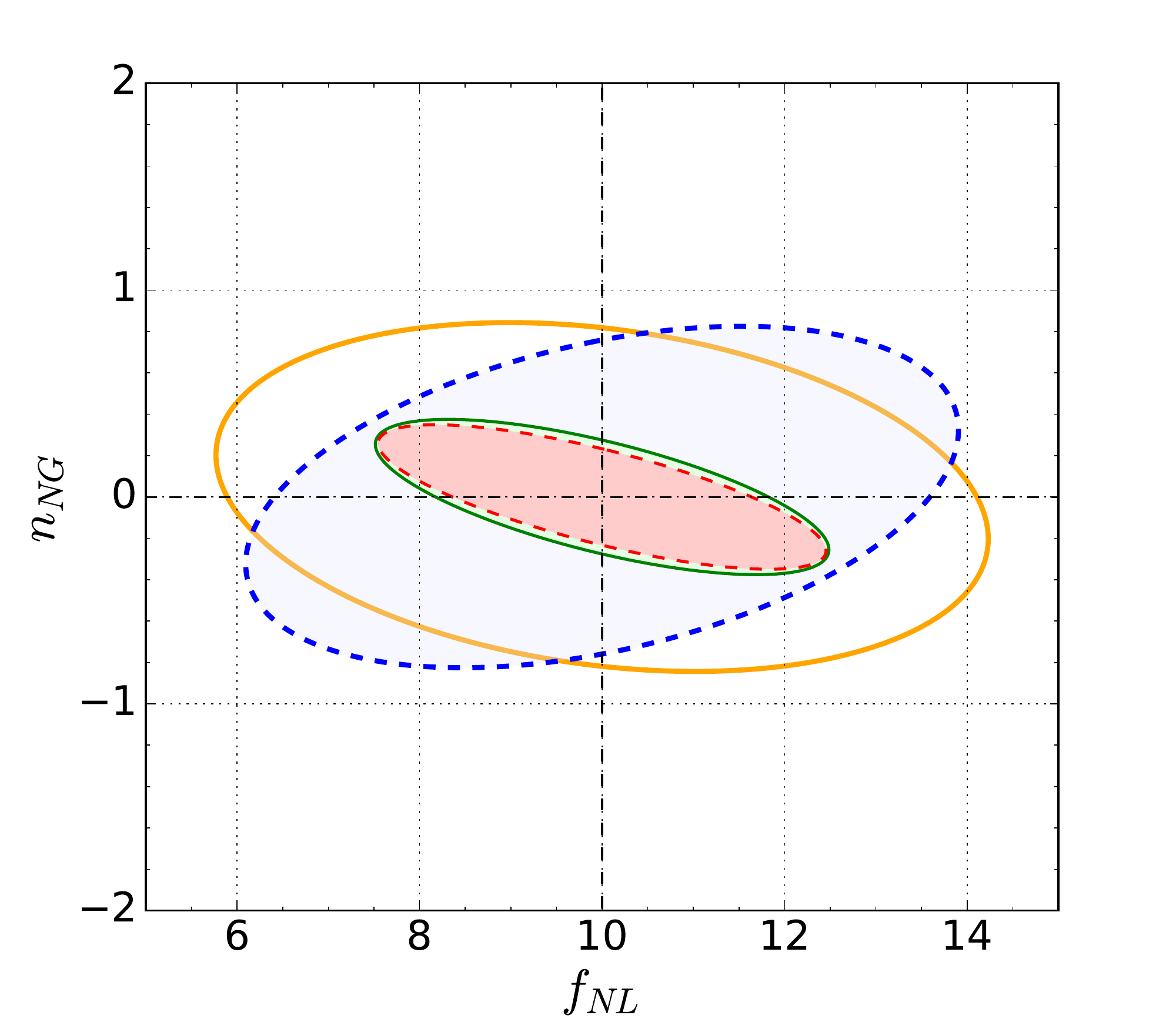}
    \includegraphics[width=0.325\textwidth]{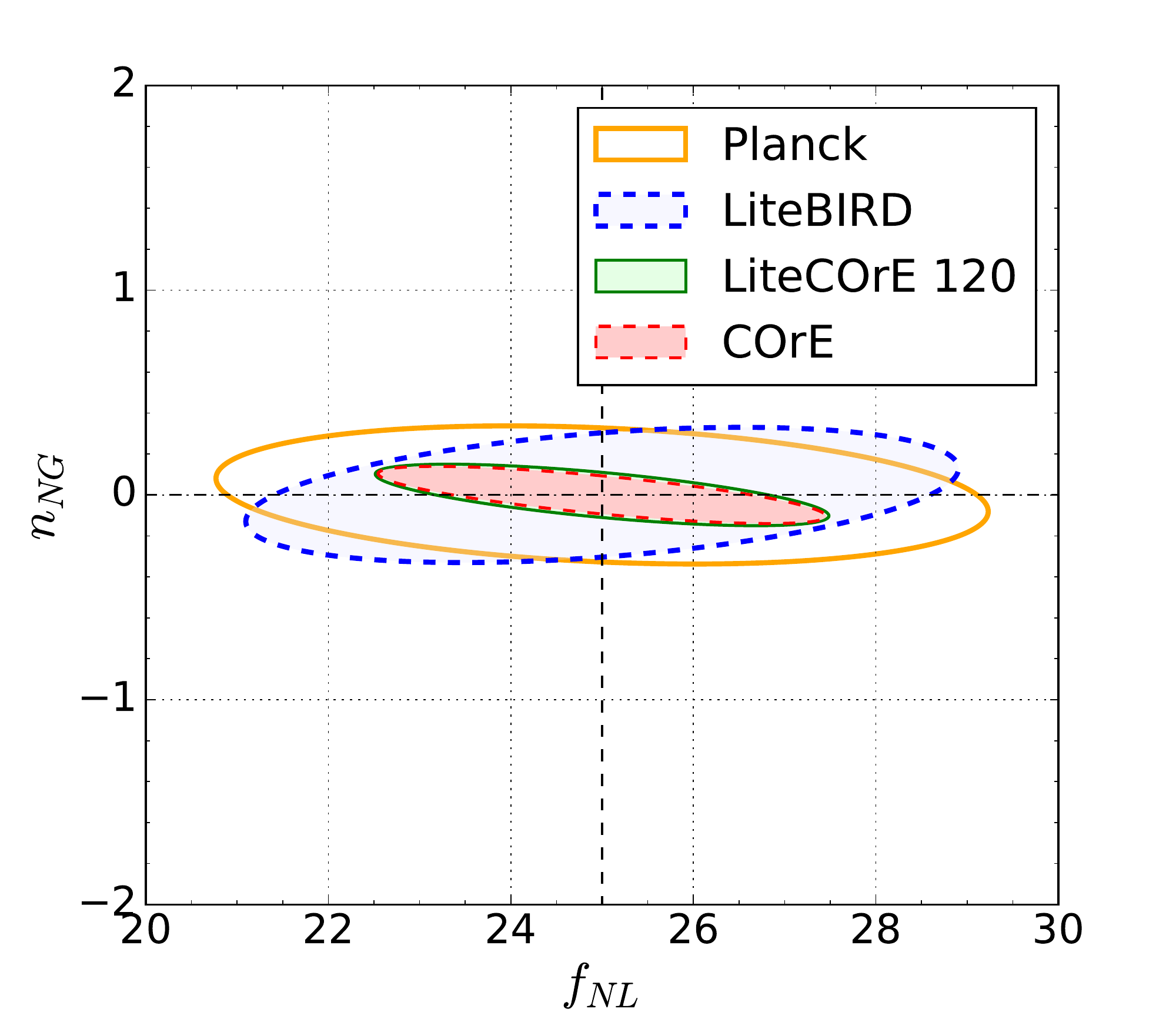}
    \caption{$1\sigma$ error ellipses for the local model arithmetic mean parametrization for different experiments, assuming joint temperature and polarization analysis and different central values of $\fnl$, from left to right:$\fnl=5,10,25$. }
    \label{fig:forsca}
\end{figure}
\begin{table}
    \centering
    \begin{tabular}{|c c|c|c|c|}
    \hline  \textbf{Experiment} & ($\ell_{max}$) & $f^{loc}_{\rm NL} = 5$ & $f^{loc}_{\rm NL} = 10$ & $f^{loc}_{\rm NL} = 25$ \\
    \hline  Planck &(2400) & $1.7$ & $0.8$ & $0.3$\\
    \hline  LiteBIRD &(1350) & $1.6$ & $0.8$ & $0.3$\\
    \hline  LiteCOrE 120 &(3000) & $0.7$ & $0.4$ & $0.1$\\
    \hline  COrE &(3000) & $0.7$ & $0.3$ & $0.1$\\
    \hline
    \end{tabular}
    \caption{Forecasts for the marginalized $1\sigma$ $n_{\rm NG}$ error bars for the arithmetic mean parametrization of the local case considering different central values for $f_{\rm NL}$. Joint temperature and polarization analysis is assumed.}
    \label{tab:forediffnl}
\end{table}
}

Constraining the running of the primordial NG parameter $f_{\rm NL}$ can provide valuable extra-information on the Physics of Inflation, allowing for better discrimination between different scenarios.

In this paper, we presented new constraints on the running of the primordial three-point function, obtained from WMAP 9-year CMB temperature data. We provided bounds on the NG running parameter $n_{\rm NG}$ for different, theory-motivated, bispectrum templates. 
Of all the models we consider (which summarize all the main running NG scenarios discussed so far in the literature), only the one-field model of equation \ref{eq:1f} was previously constrained by data \cite{PhysRevLett.109.121302}. Equilateral shape running scenarios, in particular, were not considered before in data analysis. Moreover, in previous forecasts, the model was approximated using the "geometric mean shape" of equation \ref{eq:gmpar}, rather than adopting the "arithmetic mean shape", reported in the theoretical literature on the subject \cite{Chen:2006nt}\cite{Taruya:2008pg}. We consider here the full arithmetic mean model, implementing separability via a Schwinger expansion, explicitly testing its correlation with the geometric mean ansatz and fitting it to data.

Our results were obtained by developing an estimator which extends the standard KSW bispectrum estimator, tipically used for studies of the main, scale-independent, local, equilateral and orthogonal shapes. The robustness of our pipeline was successfully tested, using a large set of scale-dependent NG simulations of the CMB sky, generated as part of this work. In our validation, we generated mock datasets including all the running bispectra under study and we considered various fiducial values of the bispectrum size and running parameters.

 Our final constraints (68\% C.L.) are $-0.6< n_{\rm NG}<1.4$ for the single-field curvaton scenario, $-0.3< n_{\rm NG}< 1.2$ for two-field curvaton case and $-1.1< n_{\rm NG} < 0.7$ for DBI. At this level of sensitivity, information coming from $n_{\rm NG}$ is not particularly useful to set new meaningful bounds on inflationary models. However, the main point of this work was to develop and test a full set of simulation and estimation tools that can be readily applied to future more sensitive datasets. 

In this respect, a natural follow-up of this study, which is actually ongoing work within the {\it Planck} collaboration, is the application of our pipeline to {\it Planck} data. Interestingly, we know that local $\fnl$, measured with {\it Planck} as a function of scale, changes from a best-fit value $\fnl \sim 40 \pm 20$, at $\ell_{\rm max} = 500$ (fully consistent with WMAP) to $\fnl \sim 2 \pm 5$ at $\ell_{\rm max} = 2000$. Fisher matrix forecasts, obtained both for {\it Planck} and other experimental setups, also show that the error bars obtained in this work could shrink up to a factor $\sim 3$ with future surveys, depending on the recovered central value of $\fnl$. Note also, see figures \ref{fig:like}, that the actual $n_{\rm NG}$ likelihood can deviate significantly from Gaussianity. This makes the predicted improvements via Fisher analysis an underestimate, for certain intervals of $\fnl$ fiducial values, which are consistent with current {\it Planck} bounds. CMB constraints, obtained via direct bispectrum estimation as presented here, could also in the future be combined with those coming from different probes, such as LSS or, in a more futuristic scenario, CMB spectral distortions and 21 cm anisotropies \cite{Emami:2015xqa,Cooray:2004kt,Biagetti:2013sr,2017JCAP...09..042R,2015JCAP...09..026K,2015JCAP...08..034R}, allowing for further, significant improvements. 

\acknowledgments

ML, NB and FO  acknowledge financial
support by ASI Grant 2016-24-H.0. NB \& ML acknowledge partial financial support by the
ASI/INAF Agreement I/072/09/0 for the Planck LFI Activity of Phase E2.
FA is supported by the National Taiwan University (NTU) under Project No. 103R4000 and by the NTU Leung Center for Cosmology and Particle Astrophysics (LeCosPA) under Project No. FI121.
We acknowledge use of the HEALPix package \cite{2005ApJ...622..759G}, \url{http://healpix.sourceforge.net}. 

\appendix

\section{KSW $f_{\rm NL}$ estimator}\label{app:KSW}
The KSW estimator exploits separability over the wavenumbers $k$ of the bispectrum shape, to achieve high numerical efficiency  \cite{0004-637X-634-1-14}. 
Starting from the assumption that the bispectrum configurations follow a Gaussian distribution, maximizing the likelihood is equivalent to minimizing the following $\chi^2$:
\begin{equation}
 \chi^2=\sum_{{\ell_1\leq\ell_2\leq\ell_3}} \frac{\left(  B_{{\ell_1\ell_2\ell_3}}^{obs} - \fnl B_{{\ell_1\ell_2\ell_3}}^{th}(
f_{\rm NL}=1)\right)^2}{\sigma^2}
\end{equation}
where $B_{{\ell_1\ell_2\ell_3}}^{th}$ and $B_{{\ell_1\ell_2\ell_3}}^{obs}$ are respectively the theoretical and observed angle averaged CMB bispectra, and $\sigma^2$ is the variance of the bispectrum, defined as:
 \begin{equation}
  \sigma_{\llll}^2=\Delta_{\llll} C_{\ell_1}C_{\ell_2}C_{\ell_3} \quad \begin{cases}
                                                           \Delta_{\llll}=1& \ell_1\neq\ell_2\neq\ell_3\\
                                                           \Delta_{\llll}=2& \ell_i=\ell_j\neq\ell_k\\
                                                           \Delta_{\llll}=6& \ell_1=\ell_2=\ell_3.\\
                                                           \end{cases}
 \end{equation}
The estimator is derived by inserting the explicit expressions for bispectra, and differentiating with respect $f_{\rm NL}$.
Using the notation introduced in \cite{2001PhRvD..63f3002K}, the angle averaged CMB bispectrum is expressed as:
\begin{equation}
B_{\ell_1 \ell_2 \ell_3} = \sqrt{\frac{(2\ell_1+1)(2\ell_2+1)(2\ell_3+1)}{4\pi}}
\begin{pmatrix} 
\ell_1 & \ell_2 & \ell_3 \\
0 & 0 & 0
\end{pmatrix}
b_{\ell_1 \ell_2 \ell_3}.
\label{eq:Bangmed1}
\end{equation}
where the matrix is the Wigner-3j symbol, encoding the geometric conditions while $b_{\ell_1 \ell_2 \ell_3}$ is the so called "reduced bispectrum", accounting for the shape dependence.
Leaving out, for the moment, the scale dependence, a general separable bispectrum template is written as as a linear combination of products of three one-dimensional functions:
\begin{equation}
B(k_1,k_2,k_3)=f_{\rm NL}\sum_{pqr} c_{pqr} F_p(k_1) F_q(k_2) F_r(k_3) + perm.
\label{eq:Sep}
\end{equation}
Projecting on the sphere, we obtain the expression for the reduced bispectrum (for the sake of clarity, but without losing in generality, here we assume a shape involving a single set of permutations):
\begin{equation}
b_{\ell_1 \ell_2 \ell_3} = f_{\rm NL}\int\mathrm{d}r \, r^2\left[\ X_{\ell_1}(r)Y_{\ell_2}(r)Z_{\ell_3}(r)+perms.\right]
\end{equation}.
Under these assumptions, the general form of the "cubic" estimator is:
 \begin{equation}
  \mathcal{E}^{cubic}=\frac{1}{\mathcal{N}}\sum_{\llll}\frac{B_{{\ell_1\ell_2\ell_3}}^{th}(f_{\rm NL}=1)B_{{\ell_1\ell_2\ell_3}}^{obs}}{C_{\ell_1}C_{\ell_2}C_{\ell_3}}=\frac{1}{\mathcal{N}}\sum_{\llll}^{\mmm}\frac{\mathcal{G}_{\ell_1 \ell_2 \ell_3}^{m_1 m_2 m_3}b_{\ell_1\ell_2\ell_3}^{\fnl=1}}{ C_{\ell_1}C_{\ell_2}C_{\ell_3}}
a\lm{1}a\lm{2}a\lm{3}
\label{eq:kswcub}
 \end{equation}
 where $\mathcal{N}$ is a normalization factor and $\mathcal{G}_{\ell_1 \ell_2 \ell_3}^{m_1 m_2 m_3}$ is the Gaunt integral, encoding the geometric properties and defined as:
\begin{align}
\mathcal{N} &=\sum_{\llll}\frac{\left(B_{{\ell_1\ell_2\ell_3}}^{th}(f_{\rm NL}=1)\right)^2}{ C_{\ell_1}C_{\ell_2}C_{\ell_3}} \label{eq:norm} \\
\mathcal{G}_{\ell_1 \ell_2 \ell_3}^{m_1 m_2 m_3} &\equiv \int \mathrm{d}^2 \hat{n} \ Y_{\ell_1}^{m_1}(\hat{n}) Y_{\ell_2}^{m_2}(\hat{n}) Y_{\ell_3}^{m_3}(\hat{n}).
\end{align}
Furthermore, in case of partial sky coverage, the rotational invariance is broken. This introduces a spurious NG signal. As a consequence the estimator becomes sub-optimal. To correct for this effect, an additional "linear" term must be added \cite{2007JCAP...03..019C}:
\begin{equation}
 \mathcal{E}^{lin}= -\frac{3}{f^{sky}\mathcal{N}}\sum_{\llll}^{\mmm}\frac{\mathcal{G}_{\ell_1 \ell_2 \ell_3}^{m_1 m_2 m_3}b_{\ell_1\ell_2\ell_3}^{\fnl=1}}{ C_{\ell_1}C_{\ell_2}C_{\ell_3}}
 \langle a_{\ell_1 m_1} a_{\ell_2 m_2} \rangle a_{\ell_3 m_3}.
 \label{eq:kswlin}
\end{equation}
where $f^{sky}$ is the fraction of the sky covered by the survey, and the brackets represent ensemble average.
The best fit value of $\fnl$ is therefore:
\begin{equation}
 \hat f_{\rm NL}= \frac{\mathcal{E}^{cubic}}{f^{sky}}+\mathcal{E}^{lin}.
\end{equation}
To obtain a numerical efficient expression, is useful to define the filtered maps:
\begin{align}
 {\mathrm M}_X(r, \hat n) =& \sum_{\ell,m} \frac{a_{\ell m}Y_{\ell}^{m}(\hat{n})}{C_\ell}X_{\ell}(r), \nonumber \\
 {\mathrm M}_Y(r, \hat n) =& \sum_{\ell,m} \frac{a_{\ell m}Y_{\ell}^{m}(\hat{n})}{C_\ell}Y_{\ell}(r), \nonumber\\
 {\mathrm M}_Z(r, \hat n) =& \sum_{\ell,m} \frac{a_{\ell m}Y_{\ell}^{m}(\hat{n})}{C_\ell}Z_{\ell}(r).
 \label{eq:filtmaps}
\end{align}
These maps can be computed resorting to fast harmonic transform.
Therefore, we can write the estimator as:
\begin{align}
 \mathcal{E}^{cubic}=&\frac{1}{\mathcal{N}}\int_0^\infty \mathrm{d}r \, r^2 \int \mathrm{d}^2 \hat{n} \, {\mathrm M}_X(r, \hat n){\mathrm M}_Y(r, \hat n){\mathrm M}_Z(r, \hat n) +perms.\\
 \mathcal{E}^{lin}=&-\frac{3}{f^{sky}\mathcal{N}}\int_0^\infty \mathrm{d}r \, r^2 \int \mathrm{d}^2 \hat{n}\,{\mathrm M}_X(r, \hat n)\langle{\mathrm M}_Y(r, \hat n){\mathrm M}_Z(r, \hat n)\rangle +perms.
\end{align}
This expression is immediately implementable and numerically efficient. This estimator has been widely used in the literature to extract the amplitude of the NG signal from CMB maps.\\
This method holds also for scale dependent templates, since the only necessary condition is bispectrum factorizability. In this case, the theoretical shape becomes a function of the parameter $n_{\rm NG}$. Fixing the value of the running, and fitting the resulting template, allows extracting the amplitude of the NG signal for a given $n_{\rm NG}$. In section \ref{sec:method}, we described how to obtain the full likelihood, including both parameters, by iterating this process over a set of $n_{\rm NG}$ values.

\section{Bispectrum shapes}
In this Appendix, we provide in details explicit expressions for the reduced bispectra considered in this paper. For an extensive review of bispectrum models and estimators, see \cite{2010AdAst2010E..73L}.

\label{ap:templates}
\subsection{Standard templates}
The reduced bispectrum for the local shape is:
\begin{equation}
b_{\ell_1\ell_2\ell_3}^{loc}=2f_{\rm NL}\int_0^\infty r^2 \mathrm{d}r\,\left[\alpha_{\ell_1}(r)\beta_{\ell_2}(r)\beta_{\ell_3}(r)+\alpha_{\ell_2}(r)\beta_{\ell_3}(r)\beta_{\ell_1}(r)+
\alpha_{\ell_3}(r)\beta_{\ell_1}(r)\beta_{\ell_2}(r)\right],
\label{eq:bloc}
\end{equation}
where:
\begin{align}
\label{eq:alpha}
\alpha_\ell(r) \equiv& \frac{2}{\pi} \int_{0}^{\infty} k^2 \mathrm{d}k \,\Delta_\ell(k,\tau) j_\ell(kr), \\
 \beta_\ell(r) \equiv& \frac{2}{\pi} \int_{0}^{\infty} k^2 \mathrm{d}k \, P_\Phi(k) \Delta_\ell(k,\tau) j_\ell(kr).
\label{eq:beta}
\end{align}
Here, $ P_\Phi(k)$ is the primordial fluctuations power spectrum and $\Delta_\ell(k,\tau)$ is the radiation transfer function. The field $\Phi$ is linked to the curvature perturbations $\zeta$ by $\Phi=\left(3/5\right)\zeta$.

The reduced bispectrum for the equilateral shape is:
\begin{equation}
b_{\ell_1\ell_2\ell_3}^{equil}=6f_{\rm NL}\int_0^\infty r^2 \mathrm{d}r\,\left[-2\delta_{\ell_1}\delta_{\ell_2}\delta_{\ell_3}-\left(\alpha_{\ell_1}\beta_{\ell_2}\beta_{\ell_3}
+2 perms.\right)+\left( \beta_{\ell_1}\gamma_{\ell_2}\delta_{\ell_3} +5 perms.\right)\right],
\label{eq:bequil}
\end{equation}
where, for simplicity of notation, we have avoided to write explicitly the $r$-dependence of quantities in the integral above. The new functions are defined as:
\begin{align}
\label{eq:gamma}
\gamma_\ell(r) \equiv& \frac{2}{\pi} \int_{0}^{\infty} k^{2}\mathrm{d}k \, P_\Phi(k)^{\frac{1}{3}}\Delta_\ell(k,\tau) j_\ell(kr), \\
\delta_\ell(r) \equiv& \frac{2}{\pi} \int_{0}^{\infty} k^2 \mathrm{d}k \, P_\Phi(k)^{\frac{2}{3}}\Delta_\ell(k,\tau) j_\ell(kr).
\label{eq:delta}
\end{align}

\subsection{Scale dependent templates}
The reduced bispectrum for the model \ref{eq:1f} is a generalization of the local shape \ref{eq:bloc}. It has the same expression and the scale dependence is encoded in the function $\alpha_\ell(r)$, that becomes:
\begin{equation}
\alpha^{1f}_\ell(r) \equiv \frac{2}{\pi}\frac{1}{\textbf{k}_{piv}^{n_{\rm NG}}} \int_{0}^{\infty} k^{2+n_{\rm NG}} \mathrm{d}k \,\Delta_\ell(k,\tau) j_\ell(kr).
\end{equation}
Similarly, the two-field model \ref{eq:2f} derives from the local template \ref{eq:bloc}, but in this case it is the $\beta_\ell(r)$ function that is promoted to a function of $n_{\rm NG}$:
\begin{equation}
\beta^{2f}_\ell(r) \equiv \frac{2}{\pi}\frac{1}{\textbf{k}_{piv}^{n_{\rm NG}/2}} \int_{0}^{\infty} k^{2+n_{\rm NG}/2} \mathrm{d}k \, P_\Phi(k) \Delta_\ell(k,\tau) j_\ell(kr).
\end{equation}
The model \ref{eq:gmpar} instead, is valid for the equilateral template \ref{eq:bequil}, the new functions being:
\begin{align}
\alpha^{gm}_\ell(r) \equiv& \frac{2}{\pi} \frac{1}{\textbf{k}_{piv}^{n_{\rm NG}/3}}\int_{0}^{\infty} k^{2+n_{\rm NG}/3}\mathrm{d}k \,\Delta_\ell(k,\tau) j_\ell(kr), \\
 \beta^{gm}_\ell(r) \equiv& \frac{2}{\pi} \frac{1}{\textbf{k}_{piv}^{n_{\rm NG}/3}}\int_{0}^{\infty} k^{2+n_{\rm NG}/3}
\mathrm{d}k \, P_\Phi(k) \Delta_\ell(k,\tau) j_\ell(kr), \\
 \gamma^{gm}_\ell(r) \equiv& \frac{2}{\pi} \frac{1}{\textbf{k}_{piv}^{n_{\rm NG}/3}}\int_{0}^{\infty} k^{2+n_{\rm NG}/3}
\mathrm{d}k \, P_\Phi(k)^{\frac{1}{3}}\Delta_\ell(k,\tau) j_\ell(kr), \\
\delta^{gm}_\ell(r) \equiv& \frac{2}{\pi} \frac{1}{\textbf{k}_{piv}^{n_{\rm NG}/3}}\int_{0}^{\infty} k^{2+n_{\rm NG}/3}\mathrm{d}k \, P_\Phi(k)^{\frac{2}{3}}\Delta_\ell(k,\tau) j_\ell(kr). 
\end{align}
Obtaining a reduced bispectrum expression for the last model \ref{eq:ampar} is slightly more complex. We have to replace $f_{\rm NL}$ with the new definition \ref{eq:scpar}, and put it in separable form, defining new coefficients.
For the local shape we compute:
\begin{align}
\label{eq:alphasd}
\alpha'_{\ell}(r,t) =&\frac{2}{\pi}\int \mathrm{d}k \,k^2\Delta_{\ell}(k)j_{\ell}(kr)e^{-tk},\\
\label{eq:betasd}
\beta'_{\ell}(r,t) =&\frac{2}{\pi}\int \mathrm{d}k \,k^2P_\Phi(k)\Delta_{\ell}(k)j_{\ell}(kr)e^{-tk},\\
\label{eq:zeta}
\zeta_{\ell}(r,t) =&\frac{2}{\pi}\int \mathrm{d}k \,k^3\Delta_{\ell}(k)j_{\ell}(kr)e^{-tk},\\
\label{eq:xi}
\xi_{\ell}(r,t) =&\frac{2}{\pi}\int \mathrm{d}k \,k^3P_\Phi(k)\Delta_{\ell}(k)j_{\ell}(kr)e^{-tk}.
\end{align}
Starting from equation \ref{eq:bloc}, with these new coefficients, we can write the reduced scale-dependent local bispectrum as:
\begin{eqnarray}
\label{eq:blocsd}
b_{\ell_1\ell_2\ell_3}^{loc} &=& \frac{f_{\rm NL}}{\textbf{k}_{piv}^{n_{\rm NG}}}\frac{1}{\Gamma(1-n_{\rm NG})}\int_0^\infty\mathrm{d}t \,
t^{-n_{\rm NG}}\int_0^\infty \mathrm{d}r \, r^2 \bigg[\Big(\alpha'_{\ell_1}(r,t)\beta'_{\ell_2}(r,t)\xi_{\ell_3}(r,t) \perm{5}\Big)+ \nonumber \\
&&+\Big( \beta'_{\ell_1}(r,t)\beta'_{\ell_2}(r,t)\zeta_{\ell_3}(r,t) \perm{2} \Big) \bigg].
\end{eqnarray}
In the equilateral case, we need to define other four additional coefficients:

\begin{eqnarray}
\label{eq:gammasd}
\gamma'_\ell(r,t) &\equiv& \frac{2}{\pi} \int_{0}^{\infty} k^2 \mathrm{d}k \, P_\Phi(k)^{\frac{1}{3}}\Delta_\ell(k,\tau) j_\ell(kr)e^{-tk}, \\
\label{eq:deltasd}
\delta'_\ell(r,t) &\equiv& \frac{2}{\pi} \int_{0}^{\infty} k^2 \mathrm{d}k \, P_\Phi(k)^{\frac{2}{3}}\Delta_\ell(k,\tau) j_\ell(kr)e^{-tk}, \\
\label{eq:epsilon}
\epsilon_\ell(r,t)  &\equiv& \frac{2}{\pi} \int_{0}^{\infty} k^3 \mathrm{d}k \, P_\Phi(k)^{\frac{1}{3}}\Delta_\ell(k,\tau) j_\ell(kr)e^{-tk}, \\
\label{eq:eta}
\eta_\ell(r,t) &\equiv& \frac{2}{\pi} \int_{0}^{\infty} k^3 \mathrm{d}k \, P_\Phi(k)^{\frac{2}{3}}\Delta_\ell(k,\tau) j_\ell(kr)e^{-tk}.
\end{eqnarray}
Inserting (\ref{eq:scpar}) in (\ref{eq:bequil}) and using all these new coefficients, we obtain the reduced equilateral bispectrum:

\begin{align}
\label{eq:bequilsd}
b_{\ell_1\ell_2\ell_3}^{equil}=&\frac{f_{\rm NL}}{\textbf{k}_{piv}^{n_{\rm NG}}}\frac{1}{\Gamma(1-n_{\rm NG})}\int_0^\infty\mathrm{d}t \,
t^{-n_{\rm NG}}\int_0^\infty \mathrm{d}r \, r^2 6\Big[-2\big( \eta_{\ell_{\ell_1}}\delta'_{\ell_2}\delta'_{\ell_3} \perm{2}\big)+\nonumber \\  
&-\big(\zeta_{\ell_1}\beta'_{\ell_2}\beta'_{\ell_3} \perm{2}\big)-\big(\alpha'_{\ell_1}\xi_{\ell_2}\beta'_{\ell_3}\perm{5} \big) +\big(\xi_{\ell_1}\gamma'_{\ell_2}\delta'_{\ell_3}\perm{5}\big)+\nonumber \\
&+\big(\beta'_{\ell_1}\epsilon_{\ell_2}\delta'_{\ell_3}\perm{5}\big)+\big(\beta'_{\ell_1}\gamma'_{\ell_2}\eta_{\ell_3}\perm{5}\big)\Big] ,
\end{align}
where we adopted a compact notation, removing the explicit dependence on $(r,t)$.

\bibliographystyle{JHEP.bst}
\bibliography{sdest.bib}

\end{document}